\documentclass[iop]{emulateapj}
\usepackage{graphicx,amsmath}

\usepackage{color}
\usepackage[plainpages=false, colorlinks=true, anchorcolor=blue, linkcolor=blue, citecolor=blue, bookmarks=false]{hyperref}

\newcommand{\Msun}{{\rm\,M_\odot}}
\newcommand{\kpc}{{\rm\,kpc}}
\newcommand{\Gyr}{{\rm\,Gyr}}
\newcommand{\kms}{{\rm\,km\,s^{-1}}}
\newcommand{\vphi}{\langle v_\phi \rangle}
\newcommand{\sigmaratio}{\sigma_{z} / \sigma_{R}}
\newcommand{\absvz}{|\langle v_z \rangle|}
\newcommand{\vz}{\langle v_z \rangle}
\newcommand{\omegadt}{d\Omega_b/dt}
\newcommand{\lzdt}{dL_{z}/dt}

\begin{document}

\title{Origin of Non-axisymmetric Features of Virgo Cluster Early-type Dwarf Galaxies. I. Bar Formation and Recurrent Buckling}

\shorttitle{Origin of Barred Dwarfs}
\shortauthors{Kwak et al.}

\author{SungWon Kwak\altaffilmark{1,2}, Woong-Tae
  Kim\altaffilmark{1,3}, Soo-Chang Rey\altaffilmark{4}, \& Suk Kim\altaffilmark{2}}

\affil{$^1$Center for the Exploration of the Origin of the Universe
  (CEOU), Astronomy Program, Department of Physics \& Astronomy,\\
  Seoul National University, Seoul 08826, Korea} %
\affil{$^2$Korea Astronomy and Space Science Institute, Daejeon 305-348, Korea} %
\affil{$^3$Center for Theoretical Physics (CTP), Seoul National University, Seoul 151-742, Korea}
\affil{$^4$Department of Astronomy and Space Science, Chungnam National University, Daejeon 305-764, Korea}

\email{swkwak@kasi.re.kr, wkim@astro.snu.ac.kr, screy@cnu.ac.kr, star4citizen@kasi.re.kr}
\slugcomment{Accepted for publication in the ApJ}

\begin{abstract}

A fraction of early-type dwarf galaxies in the Virgo cluster have a disk component and even possess disk features such as bar, lens, and spiral arms.  In this study, we construct 15 galaxy models that resemble VCC856, considered to be an infalling progenitor of disk dwarf galaxies, within observational error ranges, and use $N$-body simulations to study their long-term dynamical evolution in isolation as well as the formation of bar in them.  We find that dwarf disk galaxies readily form bars unless they have an excessively concentrated halo or a hot disk. This suggests that infalling dwarf disk galaxies are intrinsically unstable to bar formation, even without any external perturbation, accounting for a population of barred dwarf galaxies in the outskirts of the Virgo cluster. The bars form earlier and stronger in galaxies with a lower fraction of counter-streaming motions,  lower halo concentration, lower velocity anisotropy, and thinner disk. Similarly to normal disk galaxies, dwarf disk galaxies also undergo recurrent buckling instabilities. The first buckling instability tends to shorten the bar and to thicken the disk, and drives a dynamical transition in the bar pattern speed as well as mass inflow rate. In nine models, the bars regrow after the mild first buckling instability due to the efficient transfer of disk angular momentum to the halo, and are subject to recurrent buckling instabilities to turn into X-shaped bulges.
\end{abstract}


\section{Introduction}\label{s:intro}

Dwarf elliptical (dE) galaxies  are commonly known as dynamically static and morphologically elliptical stellar systems. Owing to their numerous population in galaxy clusters, they are often considered as an important key to probe the formation history of clusters. As indicated by their name, they are old and small galaxies showing no sign of recent or current star formation.
However, the discovery of two-armed spirals in VCC856 (or IC3328), a dE in the Virgo cluster, by \citet{jerjen00} has raised a controversy as to whether dEs are young and rotationally supported or old and anisotropic stellar systems.

There have since been many follow-up studies on the hidden properties of dEs \citep{barazza02, barazza03, simien02, derijcke03, geha03, lisker06a, lisker06b, lisker07, toloba11, toloba14, toloba12, janz12, janz14}.
In particular, \citet{lisker06a} performed a systematics search on 476 early-type dwarf galaxies in the Virgo cluster, finding that about one tenth of the dEs possess a disk component and that some of them even possess disk features such as spiral, bar, and lens. The fraction of the dEs with a disk component increases by more than 50$\%$ at the bright end of the dE population, and such dEs are reclassified as dEdis or dEs(di) \citep{lisker06a}. This implies that dEdi galaxies are distinct from common dEs, probably originating from genuine disk galaxies. Moreover, based on photometric studies, 15$\%$ of dEs are found with blue centers, indicating that those galaxies recently lost their gas by either ram-pressure stripping or harassment \citep{lisker06b}. In terms of the clustering properties and spatial distribution, dEdis closely follow the spatial distribution of spiral populations rather than that of common dEs  \citep{lisker07}. Based on these studies, \cite{lisker07} suggested that progenitors of dEdis are infalling, gaseous dwarf disk galaxies.

\cite{simien02} reported that some dEs, including VCC856, are found with a significant amount of rotation. Among the kinematic data of 21 dEs presented in \cite{toloba11}, the averaged rotation curve of rotationally supported dEs is similar to that of late-type galaxies at a given magnitude. By comparing the boxiness/diskiness and the anisotropic parameter, \cite{toloba11} showed that the isophotes of dEdis tend to be diskier and more rotationally supported, supporting the idea that their progenitors are disk galaxies. Using the scaling relations such as $(V-K)_{e}$-velocity dispersion and fundamental plane, \cite{toloba12} further concluded that some dEs are rather similar to late-type galaxies. Furthermore, the offset between dEs and ellipticals in the Faber-Jackson relation supports the idea that dEs are not dominated by dark matter that takes $\sim46 \pm18\%$ of the total mass within the effective radius (see also \citealt{toloba14}). By showing that 28 dEs are fast rotators among 39 Virgo dEs and contain disk-like structures, \cite{toloba15} also suggested that such dEs likely originate from late-type galaxies that lost their gas by ram-pressure stripping while infalling from the outside.

Based on the morphological features and the Sersic fitting of density profiles,
\cite{janz12,janz14} classified the Virgo dEs into four different groups, namely, one-component, two-component, population with lens, and barred population. They showed that, among the dEdis, the barred population is $\sim35 \pm9 \%$ while the lens population is $\sim8 \pm5 \%$. Lenses are known as `defunct bars' \citep{kormendy79,kormendy04} because they have density profiles similar to bars, but shorter major axis. Although these barred populations take nearly a half of the entire Virgo dEdis, their origin remains mysterious. Observations indicate that there is a considerable number of barred dEdis at the outskirts of the Virgo cluster \citep{janz12, janz14}. Using $N$-body simulations, \cite{lokas16} recently showed that the tidal potential in a Virgo-like cluster can trigger the bar formation in massive disk galaxies after the pericenter approach. However, since the orbits are unknown, it is not certain whether observed barred dEdis at the cluster outskirts have been subject to stronger tidal perturbations. The main purpose of the present paper is to show that the bar formation in dEdis does not require a tidal forcing: they are intrinsically unstable to forming bars.

The bar formation in disk galaxies is usually explained by dynamical instabilities of self-gravitating, axisymmetric stellar disks \citep{miller70, hohl71, kalnajs72, kalnajs77}. There are certainly many factors that affect the stability of disk galaxies, including the mass of dark matter halo
\citep{ostriker73, christodoulou95, sellwood01}, a fraction of counter-rotating stars in the disk \citep{sellwood94}, and disk thickness \citep{klypin09}.
For example, it is well known that the dark halo plays a dual role: a massive spheroidal component alleviates disks against forming bar-like structures \citep{ostriker73}, while a responsive halo destabilizes the system by allowing angular momentum transfer from the disk to halo and thus expedites the bar growth \citep{athanassoula03}. In addition, $\sim 20-50\%$ of the total disk mass is frequently observed to be in the counter rotation  \citep{bettoni90,merrifield94,bertola99,kannappan01a}, possibly caused in numerical simulations by gas infall with opposite spins and/or minor mergers \citep{thakar96, thakar98, algorry14}. These counter-rotating stellar components tend to stabilize the disk by making its center hotter \citep{sellwood94}, as well as by lowering its rotational kinetic energy \citep{ostriker73,efstathiou82}. Since most of the previous studies mentioned above considered normal disk galaxies, in this work we focus on dwarf disk galaxies, mimicking VCC856 in the Virgo cluster, and study how galaxy properties change the intrinsic stability of dEdis against the bar formation.

Once a bar forms, it undergoes another transition in the vertical direction called buckling instability. It partially weakens and shortens the bar by disrupting its outer parts \citep{combes81, combes90, raha91, merritt94, martinez04}. Observationally, \cite{erwin16} caught the buckling instability in action in two nearby galaxies. \cite{berentzen06} emphasized that adopting a live halo is crucial to exploring the buckling instability. The buckling instability is also known as a formation channel of a boxy/peanut/X-shaped bulge \citep{combes90, raha91, martinez04, berentzen06, athanassoula16}.
The local theory of \cite{toomre66} and \cite{araki85} suggests that the buckling instability occurs when the ratio $\sigma_z/\sigma_R$  exceeds $0.3$ when the disk is uniform, where $\sigma_z$ and $\sigma_R$ denote the velocity dispersions in the vertical and radial directions, respectively. However, the ratio varies with radius in real galaxies, so that one single threshold cannot determine the occurrence of buckling instability \citep{raha91}. Interestingly, \cite{martinez06} reported that the buckling instability is recurrent due to the regrowth of bars. Multiple renewals of bars with repeated destruction and formation helped by gas accretion were investigated by \citet{bournaud02}, but the condition for regrowth in isolation is not well understood yet.

In this paper, we run $N$-body simulations of dwarf disk galaxies to study their long-term dynamical evolution in isolation.  We construct galaxy models that resemble VCC865, which is thought to be an infalling progenitor of dEdis.  The primary purposes of this work are to understand what it is that forms bars in dEdis and to see how differently dEdis evolve as compared to normal disk galaxies.  We also vary galaxy properties within observational error ranges to explore the effects of the counter streaming motion, velocity dispersion, disk scale height, dark matter fraction, and stellar disk mass on the bar formation and recurrent buckling instability occurring in isolated dwarf disk galaxies.

The organization of this paper is as follows. In section \ref{s:ch2}, we describe our galaxy models and numerical methods. In Section \ref{s:ch3}, we present the detailed evolution of our fiducial model in terms of the size, strength, and pattern speed of a bar that forms. In section \ref{s:ch4}, we explore the effects of the various parameters on the bar formation and buckling instability. In section \ref{s:ch5}, we summarize and discuss our results in an astronomical context.

\section{Models and Method}\label{s:ch2}

In this section, we describe our galaxy models and simulation methods. We first justify that VCC856 can be a progenitor galaxy in view of its kinematics and morphology. We then explain our initial conditions and model parameters as well as the numerical methods we adopt.

\subsection{VCC856}

As an initial condition of our simulations, it is ideal to choose a dwarf disk galaxy that maximally represents the infalling population without undergoing environmental effects. Among numerous dEdis in the Virgo cluster, we take VCC856 as our fiducial progenitor galaxy for the following reasons. The morphological classification of VCC856 was a nucleated dE in \cite{binggeli85}, but \cite{jerjen00} first reported the presence of faint logarithmic spiral arms after subtracting mean axisymmetric light distribution.
The pitch angle of the spirals is $12.^\circ1$, similar to those in Sb or Sbc galaxies. For this disk feature, VCC856 was reclassified as a dEdi or dE(di) by \cite{lisker06b}, and its properties have been inspected observationally  in the literature \citep{barazza02, simien02,  geha03, jerjen04, chilingarian09,lisker09, toloba11, kormendy12}.

The photometric data show that the surface brightness profile of VCC856 follows an exponential distribution without a bulge \citep{jerjen00,lisker09,janz14,toloba14}, implying that it has not yet experienced any major mergers. If a bulgeless dwarf disk galaxy has experienced either strong tidal effects or mergers, it would probably have a bar or non-exponential density profile. In addition, unlike early-type galaxies that are commonly pressure-supported, VCC856 is a rotationally supported system \citep{jerjen00,simien02,lisker09,toloba11,toloba14}, which strengthens our assumption that VCC856 is a genuine dwarf disk galaxy. Finally, it is a nearly gas free galaxy and lacks prominent \ion{H}{2} regions \citep{jerjen00}, suggesting that it has entered the Virgo Cluster relatively recently and has experienced ram-pressure stripping without considerable influence on its stellar disk. Although it already possesses spiral arms  presumably driven by a weak tidal encounter, the fractional arm amplitude is only $\sim3-4 \% $ \citep{jerjen00}. Therefore, we regard VCC856 as an infalling progenitor that preserves the information of the early phase of the infalling dwarf disk galaxies. VCC886 is one of the best studied dEdis with sufficient information available for numerical modeling.

\subsection{Initial Conditions}

Our galaxy models, mimicking VCC856, consist of a stellar disk and a live halo.
Although the presence of a nucleus in VCC856 is apparent (e.g., \citealt{jerjen00,lisker09}), we do not include it in our models since its size is too small to affect global dynamics of the stellar disk.
For the dark halo, we adopt the \cite{hernquist90} profile
\begin{equation}\label{eq:hernquist}
\rho_\text{DM} (r) = \frac{M_\text{DM}}{2\pi} \frac{a}{r(r+a)^3} ,
\end{equation}
where $r$ is the radial distance, $M_\text{DM}$ is the total mass, and $a$ is the scale radius of the halo. The scale radius is related to the concentration parameter $c$ as
\begin{equation}\label{eq:cc}
a = \frac{r_{200}}{c} \left[2 \ln(1+c)-\frac{c}{1+c}\right]^{1/2},
\end{equation}
with $r_{200}$ being the virial radius  \citep{springel05}. For all models, we fix $r_{200} = 40$ kpc and  $M_\text{DM} = 1.5  \times 10^{10} \Msun$, while varying $c$ (see below).

The initial density distribution of the stellar disk is
\begin{equation}\label{eq:expdisk}
\rho_{\star} (R, z) = \frac{M_{d}}{4\pi z_{d} R_{d}} \exp \left( -\frac{R}{R_{d}} \right)  \text{sech}^2 \left(\frac{z}{z_{d}}\right),
\end{equation}
where $R$ and $z$ denote the radial and vertical distances in the cylindrical coordinates, respectively, $M_{d}$ is the total mass of the disk, and $z_{d}$ is the vertical scale height. The corresponding surface density is
$\Sigma_{\star} = \int_{-\infty}^\infty \rho_{\star} dz = M_{d} \exp(-R/R_{d}) /2\pi R_d $. The disk scale radius $R_{d}$ is determined by taking allowance for the spin parameter, as described in \cite{mo98}. The effective radius that encloses a half of the stellar mass is $R_\text{eff}=1.68R_d$. For most cases, we take $z_d=0.33R_d$ obtained from \cite{lisker09} that is identical to the median axial ratio of the dEdis from \cite{lisker07}. In some models we vary it to study the effect of disk thickness on the bar formation.

The observed stellar mass in VCC856 is $\log M_{e}^{\star} = 8.72 \pm 0.12 \Msun$ where $M_{e}^{\star}$ is the measured value within the effective radius \citep{toloba14}, so we take $M_d=10^9\Msun$ as the total disk mass in our standard model S1. This corresponds to $\sim7\%$ of the total galaxy mass.
The relative importance of dark matter near the central parts is measured by the dark matter fraction \emph{inside} the effective radius
 \begin{equation}
  f_\text{DM} = \left( 1 + \frac{2\int_0^{R_\text{eff}} \rho_\text{DM} r^2 dr}{\int_0^{R_\text{eff}} \Sigma_\star r dr}\right)^{-1},
 \end{equation}
which can be controlled by the halo concentration parameter $c$.
For model S1 with $c =9$, $R_\text{eff}=1.05\kpc$, and $f_\text{DM}=30\%$.
This is very similar to the measured value $f_\text{DM}=0.33 \pm 0.26$ of VCC856 \citep{toloba14}.

For the comparison with $f_{\rm{DM}}$ in large disk galaxies, we briefly construct Milky Way-like galaxies by varying the halo mass and concentration obtained from the Aquarius Project \citep{springel08}. We find that the values of $f_{\rm DM}$ range between $54$--$60\%$ within $R_{\rm eff}$.

\subsection{Galaxy Models}

For a given mass distribution, the vertical velocity dispersion $\sigma_z$ is automatically constrained by $z_d$. Our standard model has $\sigma_z\sim25\kms$ at the center, which is consistent with the observed central velocity dispersion $\sigma_{c}\sim 27\pm4 \kms$ of VCC856 \citep{peterson93}. To adjust the radial velocity dispersion $\sigma_R$, we use the velocity anisotropy parameter
\begin{equation}\label{eq:disp}
f_{R} = \frac{\sigma_{R}^2}{\sigma_{z}^2},
\end{equation}
\citep{yurin14}.
Following \citet{gerssen00} and \citet{lisker09}, we adopt $f_R=1.56$ applicable to VCC856. This is smaller than $f_R\sim 4 $ of the Milky Way in the solar neighborhood that is much thinner than dEdis and might have experienced radial heating due to the central bar.

\begin{figure}
\centering\includegraphics[angle=0,width=8.5cm]{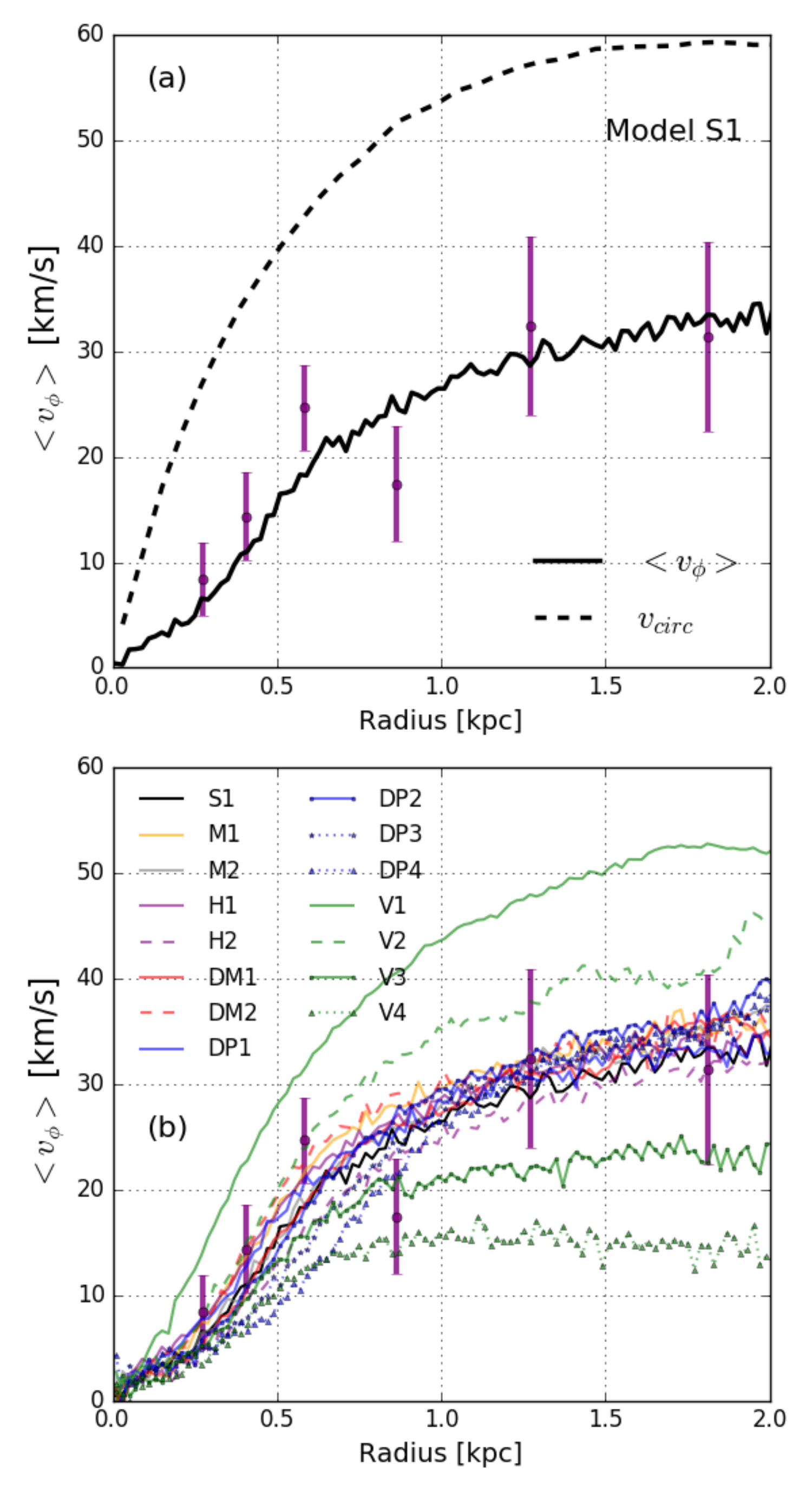}
\caption
{Mean azimuthal velocity $\vphi$ (lines) compared to the rotation curve of VCC856 corrected for an inclination angle (dots with error bars) for (a) the standard model S1 and (b) all models. The dashed line in (a) draws the circular velocity $v_{\rm circ}$, obtained by computing the disk potential in the midplane.}\label{fig01dynamics}
\end{figure}

With the stellar mass and the halo concentration specified above, the mean azimuthal velocity $\vphi$ readily exceeds observed rotational velocities of dEdis if all stellar particles rotate in the same sense. One can reduce $\vphi$ by introducing counter rotating particles in the azimuthal direction, the amount of which can be parameterized as
\begin{equation}\label{eq:k}
k^2 \equiv \frac{\vphi^{2}}{\langle v^2_{\phi} \rangle-\sigma^2_{R}},
\end{equation}
where the angle brackets $\langle \;\rangle$ denote the average over the azimuthal direction \citep{satoh80,yurin14}.\footnote{The axisymmetric Jeans equations are independent of $\vphi$, so that $k^2$ does not change the density structure of an equilibrium configuration.} Our standard model S1 has $k = 0.65$. As shown in Figure \ref{fig01dynamics}(a), this yields the rotation curve similar to the observed rotation curve of  VCC856 \citep{simien02}
after correcting for the inclination angle of $25.^\circ2$ \citep{jerjen00, simien02}.

\begin{deluxetable}{cclllll}
\tablecaption{Model Parameters\label{tbl:models}}
\tablewidth{0pt}
\tablehead{\colhead{Model}
          & \colhead{$M_{d}$}
          & \colhead{$z_{d}$}
          & \colhead{$f_\text{DM}$}
          & \colhead{$f_R^{1/2}$}
          & \colhead{$k$} \\
            \colhead{$~$}
          & \colhead{[$10^9 \Msun$]}
          & \colhead{$[R_{d}]$}
          & \colhead{$\%$ ($c$)}
          & \colhead{$~$}
          &  \colhead{$~$} \\
            \colhead{(1)}
          & \colhead{(2)}
          & \colhead{(3)}
          & \colhead{(4)}
          & \colhead{(5)}
          & \colhead{(6)} }
\startdata
S1  & 1.0 & 0.33 & $\sim30\%$ (9)  & 1.25 & 0.65 \\[1ex]
V1  & 1.0 & 0.33 & $\sim30\%$ (9)  & 1.25 & \textbf{1.0}   \\
V2  & 1.0 & 0.33 & $\sim30\%$ (9)  & 1.25 & \textbf{0.825} \\
V3  & 1.0 & 0.33 & $\sim30\%$ (9)  & 1.25 & \textbf{0.475} \\
V4  & 1.0 & 0.33 & $\sim30\%$ (9)  & 1.25 & \textbf{0.3}   \\[1ex]
DP1 & 1.0 & 0.33 & $\sim30\%$ (9)  & \textbf{1.0}  & \textbf{0.65}  \\
DP2 & 1.0 & 0.33 & $\sim30\%$ (9)  & \textbf{1.5}  & \textbf{0.75}  \\
DP3 & 1.0 & 0.33 & $\sim30\%$ (9)  & \textbf{1.75} & \textbf{0.75}  \\
DP4 & 1.0 & 0.33 & $\sim30\%$ (9)  & \textbf{2.0}  & \textbf{0.8}   \\[1ex]
H1  & 1.0 & \textbf{0.23} & $\sim30\%$ (9)  & 1.25 & 0.65  \\
H2  & 1.0 & \textbf{0.43} & $\sim30\%$ (9)  & 1.25 & 0.65  \\[1ex]
DM1 & 1.0 & 0.33 & $\sim\textbf{7\%}$ \textbf{(2)}  & 1.25 & \textbf{0.8}   \\
DM2 & 1.0 & 0.33 & $\sim\textbf{54\%}$ \textbf{(20)} & 1.25 & \textbf{0.55}   \\[1ex]
M1  & \textbf{1.4} & 0.33 & \textbf{$\sim30\%$ (12)} & 1.25 & \textbf{0.6}   \\
M2  & \textbf{0.8} & 0.33 & \textbf{$\sim32\%$ (8)}  & 1.25 & \textbf{0.75}  \\[1ex]
\enddata
\tablecomments{Column (1) is the model name. Columns (2) and (3) give  the total disk mass in units of $10^9\Msun$ and the disk scale height relative to $R_d$, respectively. Column (4) gives the fraction of dark matter within the effective radius. Column (5) lists the ratio of the velocity dispersions. Column (6) lists the mean streaming parameter. The variations from Model S1 are indicated in boldface.}
\end{deluxetable}

We construct a total of 15 models that reflect the error ranges in the dynamical properties of VCC856. Table \ref{tbl:models} lists the name and parameters of our models. Relative to model S1, the V models differ only in the value of $k$ that changes $\vphi$ by controlling the fraction of counter rotating stars, as Figure \ref{fig01dynamics}(b) shows. The DP models explore the effects of $f_{R}$ on the bar formation. In the H models, we vary the scale height by $0.1R_{d}$. In DM models, we control the halo concentration factor $c$ to change $f_\text{DM}$ within the effective radius. The M models vary the total disk mass, while keeping $f_\text{DM}$ the same as in model S1. By fixing $z_d$, the M models have different $\sigma_z$, although the ratio $\sigma_R/\sigma_z$ remains similar to the one in model S1. Except for the V models, we change the $k$ parameter as well in all the models to make the mean streaming velocity identical to the one in model S1.

\begin{figure}
\centering\includegraphics[angle=0,width=8.5cm]{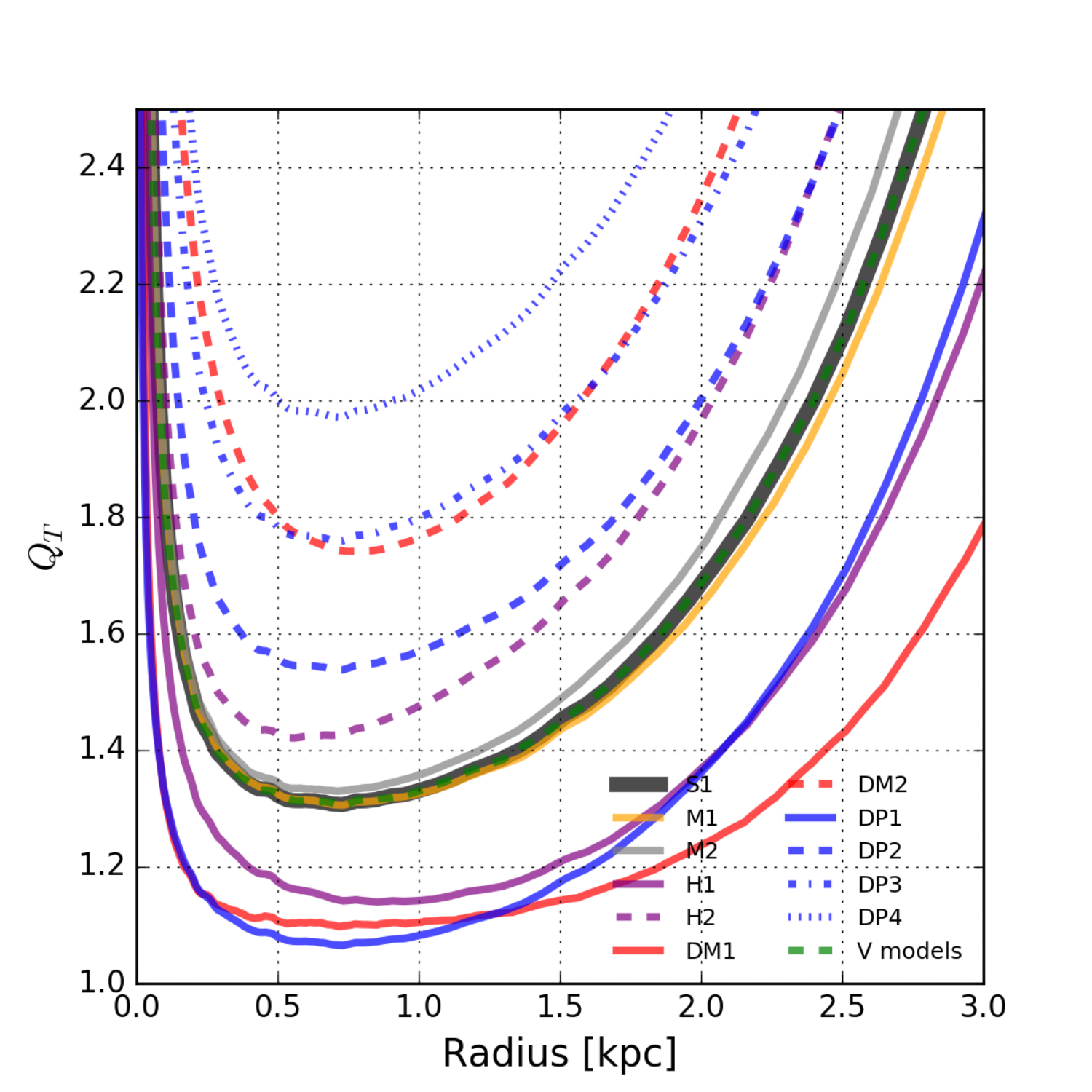}
\caption
{Comparison of the Toomre stability parameter $Q_T$ of the initial disks in all the models.}\label{fig02toomre}
\end{figure}

Figure \ref{fig02toomre} plots the radial distributions of the \citet{toomre64} stability parameter
\begin{equation}\label{eq:toomre}
Q_T = \frac{\kappa \sigma_{R}}{3.36 G \Sigma_\star},
\end{equation}
measuring local susceptibility of an initial stellar disk to gravitational perturbations. Here, $\kappa$ is the epicycle frequency and $G$ is the gravitational constant. Our standard model has $Q_T\approx1.3$ at $R=0.5-1$ kpc. Models H1, DP1, and DM1 are more unstable than our standard model, while the M and V models have nearly the same $Q_T$ as model S1. However, we note that $Q_T$ does not take into account the random kinetic energy in the azimuthal and vertical directions, which may play an important role in the bar formation \citep{ostriker73, efstathiou82, klypin09}.

\subsection{Methods}
We use the publicly available \textsc{galic} code \citep{yurin14} to construct the initial configurations of model galaxies. For a given density distribution, \textsc{galic} iteratively adjusts particle velocities to achieve an equilibrium configuration. The flexibility of \textsc{galic} in generating outcomes, such as the mean streaming velocity and the ratio of velocity dispersions in the radial and vertical directions, allows us to easily obtain our desired initial conditions.

Each galaxy model is constructed by distributing a total of $2.5 \times 10^{6}$ particles, while keeping the mass ratio of a dark matter particle to a stellar particle to 3.  Model S1 and other models with the same disk mass thus employ $5 \times 10^{5}$ particles for the stellar disk and $2 \times 10^{6}$ for the dark halo. For the softening lengths of the stellar and dark matter particles, we adopt the mean particle separations inside the half-mass radius. All our simulations are performed using the \textsc{changa} code \citep{jetley08, jetley10, menon15}, an $N$-body \textsc{treecode} + SPH code. We set the force accuracy option and the timestepping scale to $\theta=0.7$ and $\eta=0.1$.

For data analysis, we use \textsc{pynbody}, which is a python package for analyzing outputs of $N$-body and SPH simulations \citep{pynbody}. Since all initial conditions are generated in the \textsc{gadget} format from the \textsc{galic} code, the units of mass, length, and velocity are $10^{10} \Msun$, $1\kpc$, and $1\kms$, respectively. The time unit is 0.9785 Gyr. All the simulations are run for 10 Gyr.

\section{Evolution of the Standard Model}\label{s:ch3}

In this section, we present the evolution of model S1 to study the bar formation in typical dEdis. We focus on the temporal variations of the bar length and strength, as well as buckling instability occurring in model S1. The effects of changing galaxy parameters will be presented in Section \ref{s:ch4}.

\subsection{Overall Evolution}

\begin{figure*}
\centering\includegraphics[angle=0,width=17cm]{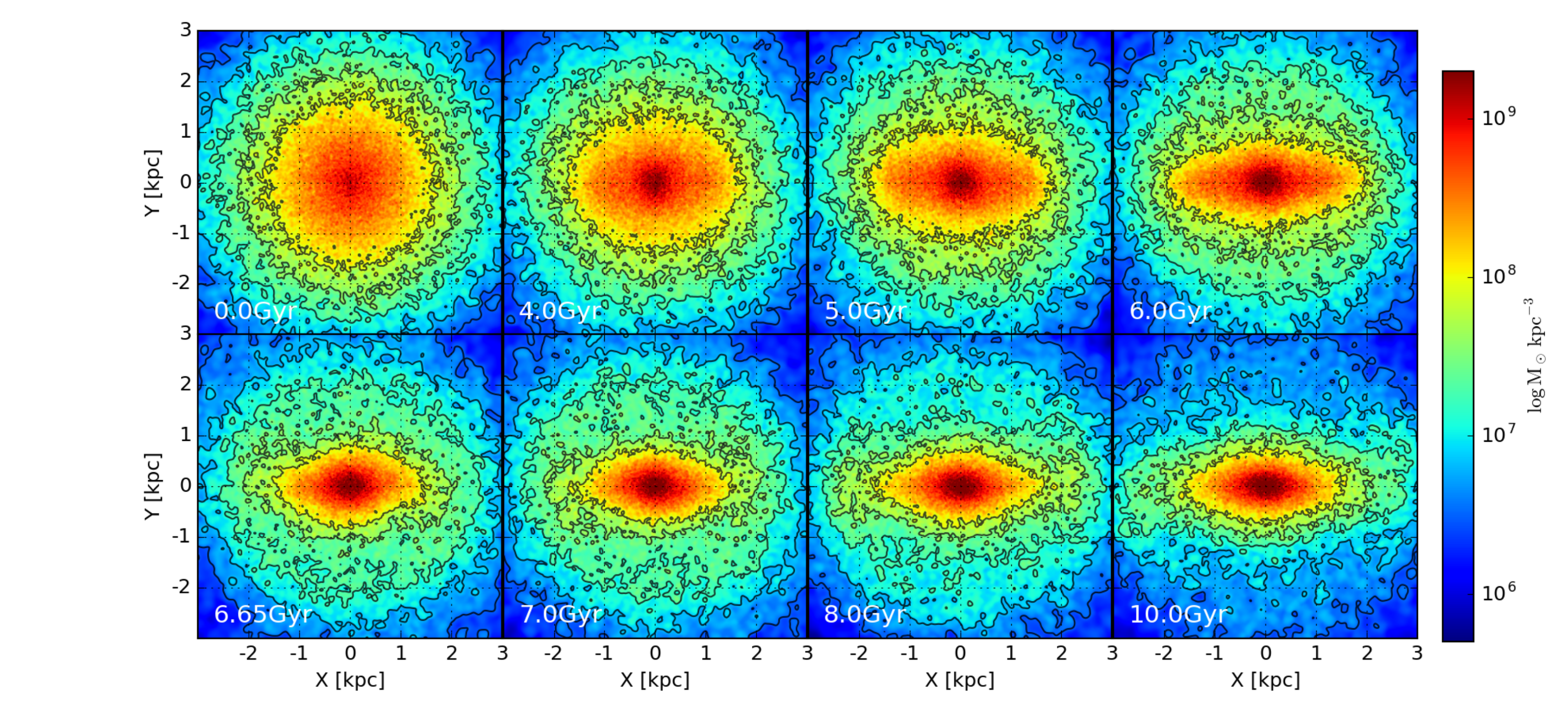}
\centering\includegraphics[angle=0,width=17cm]{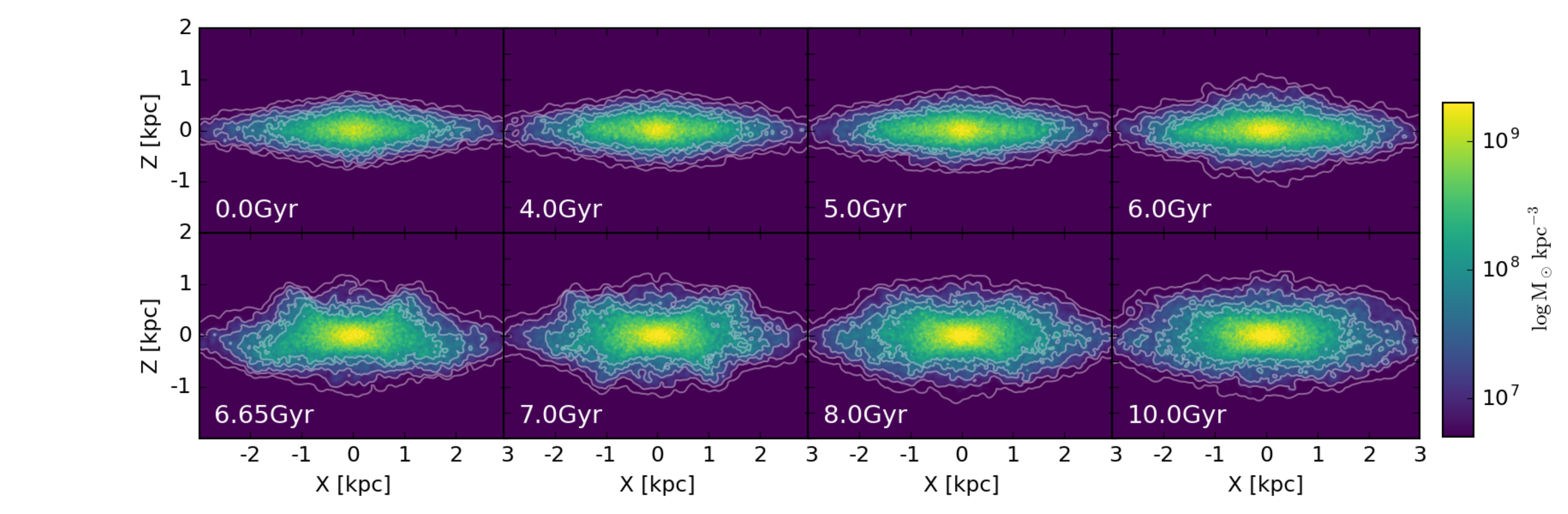}
\vspace{-0.2cm}
\caption
{Face-on (upper panels) and edge-on (lower panels) snapshots of the disk density in our standard model S1.
The corresponding time is labeled in the lower left corner of each panel. Contours are drawn in logarithmic scale between $10^{7.89}$ and $10^{6.50} {\rm\,M_\odot}{\rm\,kpc}^{-3}$. The highest contour level corresponds to the bar boundary.
See the text for further detail.
}\label{fig03image}
\end{figure*}

Figure \ref{fig03image} displays the snapshots of the disk density in model S1 at the $z=0$ (upper panels) and $y=0$ (lower panels) planes. A bar begins to form at $t=2\Gyr$ as a result of global disk instability. The bar becomes thinner and longer as it grows, eventually becoming very prominent at around $t=6\Gyr$. Subsequently, it experiences a buckling instability, which not only shortens the bar but also causes it to bend asymmetrically out of the $z=0$ plane at $R=1$--$2$ kpc, as shown by the $t=6.65$ and $7\Gyr$ snapshots. At $t=8\Gyr$, the bar regrows slightly longer and becomes more-or-less symmetric relative to the $z=0$ plane. It then undergoes a second episode of buckling instability, which  effectively thickens the disk again at larger radii ($R=2$--$3$ kpc) than the first one. Since the second buckling is quite weak, it leads to an X-shaped bulge at $t=10\Gyr$, less pronounced compared to those in normal barred galaxies (e.g., \citealt{martinez06}).
As a result of the buckling instabilities, the axial ratio of the stellar disk increases by a factor of about 2.

\begin{figure}
\centering\centering\includegraphics[angle=0,width=8.5cm]{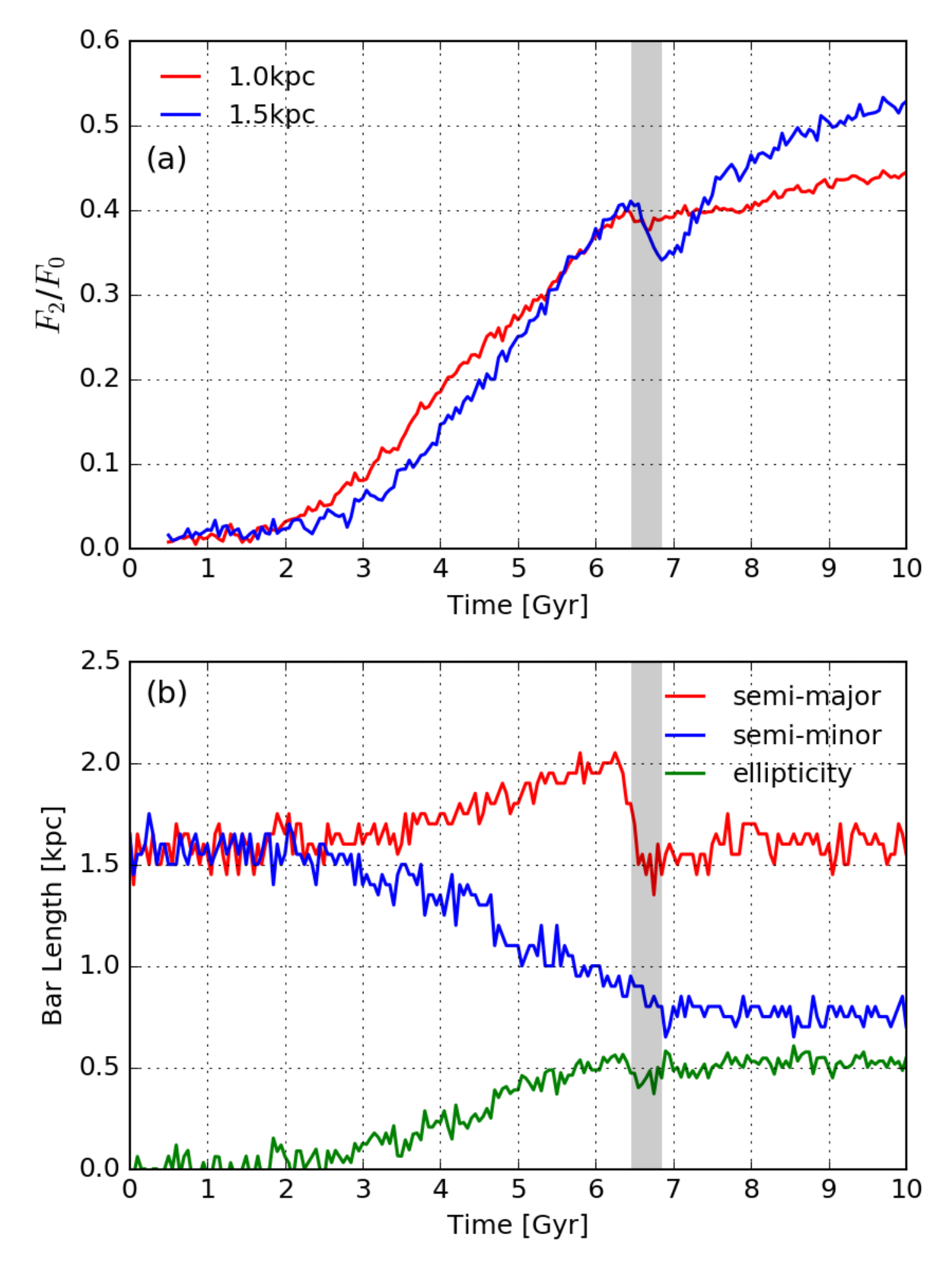}
\caption
{Temporal changes of the bar strength (top) and the bar lengths and ellipticity (bottom) in model S1. The shaded vertical strips between $t=6.4$ and 6.9 Gyr mark the local maximum and minimum of the bar strength at $R=1.5$ kpc, corresponding to the occurrence of the first buckling instability.}\label{fig04barstrength}
\end{figure}

We quantify the bar strength at each time by measuring the ratio $F_{2} / F_{0}$, where $F_m$ denotes the Fourier transform
\begin{equation}\label{eq:fdm}
   F_{m} (R) = \sum_{j} \mu_{j} e^{im\phi_j},
\end{equation}
with $\mu_j$ and $\phi_j$ being the mass and azimuthal angle of the $j$-th particle in an annulus with width $\Delta R=0.1\kpc$ centered at $R$. Figure \ref{fig04barstrength}(a) plots the ratio at $R=1.0$ and $1.5\kpc$ using the red and blue lines, respectively. It is apparent that the bar starts to grow from $t\sim2\Gyr$, achieves the maximum strength at $t\sim6.4\Gyr$ right before undergoing the first buckling instability that temporarily weakens the bar. Note that the bar strength at $R=1.5\kpc$ increases again after the buckling instability, while the buckling instability appears quite ineffective inside $R=1.0\kpc$.

Since the bar is embedded in the disk and its density continuously varies, it is non-trivial to determine the bar semi-major/minor axes. We empirically find that $\rho=10^{7.89} {\Msun}{\kpc}^{-3}$, corresponding to the initial disk density at $(R,z)=(1.25R_{d},0)$, reasonably well describes the boundaries of the bars formed in our simulations.  Figure \ref{fig04barstrength}(b) plots the temporal variations of the semi-major axis $a$ and the semi-minor axis $b$, as well as the bar ellipticity $\epsilon=1-b/a$. The semi-major axis begins to deviate from the initial value as the bar grows and reaches its maximum value of $\sim2\kpc$ at $t=6.3\Gyr$. After the buckling instability, the semi-major axis is shortened rapidly to $\sim1.5\kpc$ followed by a small, gradual increment with some fluctuations, while the bar strength at $R=1.5\kpc$ tends to increase continuously.  As evidenced by the gradual deformation of low-density contours in the face-on views at $t=7-10$ Gyr in Figure \ref{fig03image}, the bar regrowth in the post-buckling period is caused primarily by the mass accumulation in the regions near the bar ends.\footnote{The increase in the bar length is only slight in Figure \ref{fig04barstrength}(b) since the bar is defined as the high-density regions with $\rho\geq 10^{7.89} {\Msun}{\kpc}^{-3}$.}

\begin{figure}
\centering\includegraphics[angle=0,width=8.5cm]{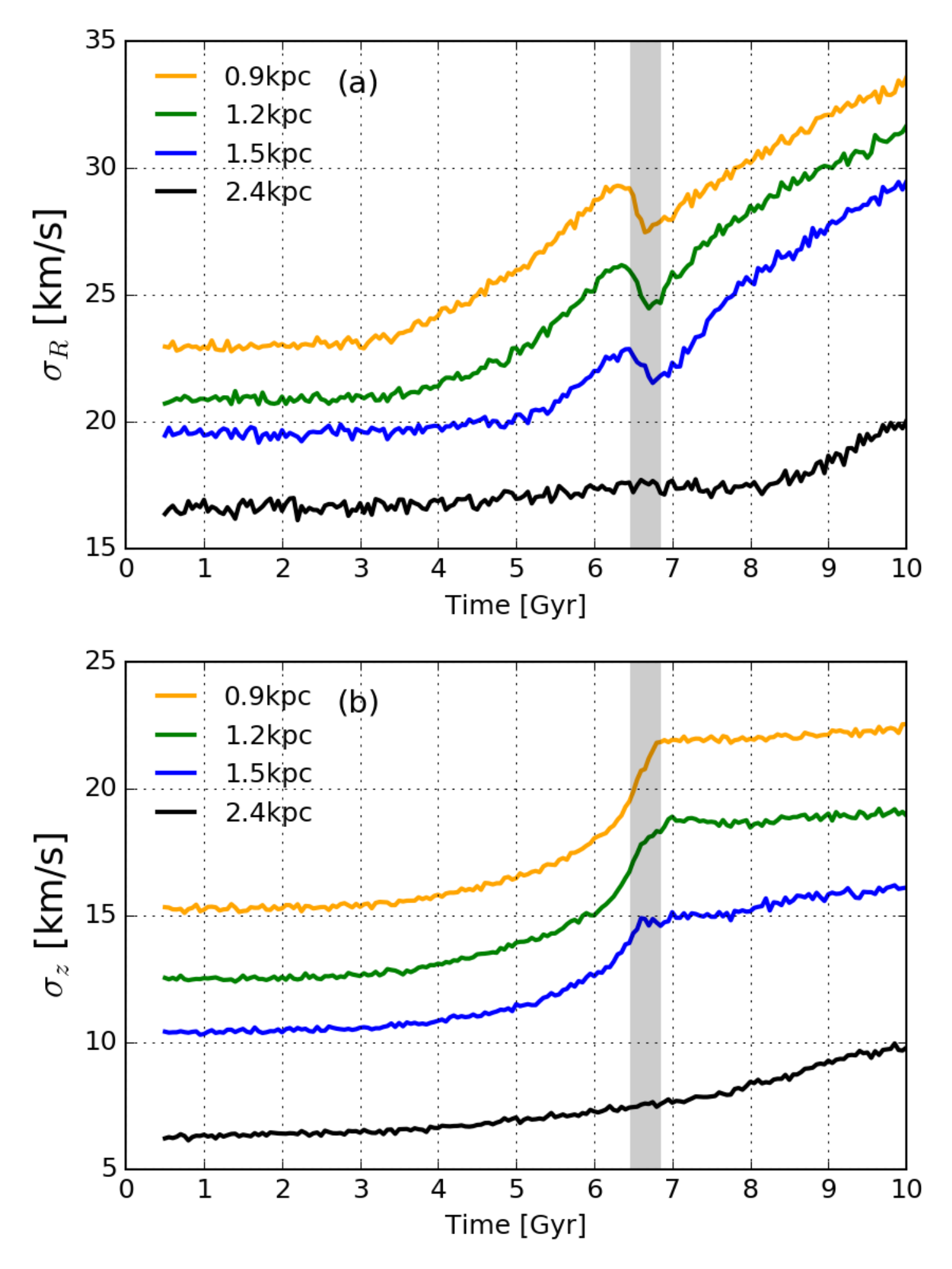}
\caption
{Evolution of the radial velocity dispersion $\sigma_R$ (top) and the vertical velocity dispersion $\sigma_z$ (bottom) in model S1.
The shaded vertical strips  between $t=6.4$ and 6.9 Gyr mark the occurrence of the buckling instability.}\label{fig05disp}
\end{figure}

Figure \ref{fig05disp} plots the evolution of the radial and vertical velocity dispersions of the disk in model S1 at four different radii.  The bar formation induces non-circular motions of the disk particles, gradually increasing $\sigma_R$.  The increase of $\sigma_R$ occurs earlier at smaller $R$, indicating that the bar begins to shape from the inner disk.  The region at $R=1.5\kpc$ starts to respond only after $t=5\Gyr$ when the bar becomes sufficiently long to affect the outer regions, and the region at $R=2.4\kpc$ outside the bar remains almost unaffected by the bar formation. The buckling instability occurring at $t=6.4\Gyr$ makes $\sigma_R$ reduced temporarily in the bar regions. The subsequent bar growth increases $\sigma_R$ in the bar regions. At $R=2.4\kpc$ outside the bar, $\sigma_R$ visibly increases after $t=8\Gyr$ due to the bar regrowth.

The vertical velocity dispersion also increases with time as the bar grows, but more mildly than $\sigma_R$. A relatively rapid increase of $\sigma_z$ is observed near $t=6.4\Gyr$ due to the first buckling which triggers vertical heating by dissolving the outer parts of the bar \citep{combes81, combes90, raha91, merritt94, martinez04}. Unlike $\sigma_R$, the vertical velocity dispersion at $R\lesssim 1.2\kpc$ stays almost constant after the buckling instability is over, although it increases slightly faster  after the first buckling instability outside the bar regions.

\begin{figure}
\centering\includegraphics[angle=0,width=8.5cm]{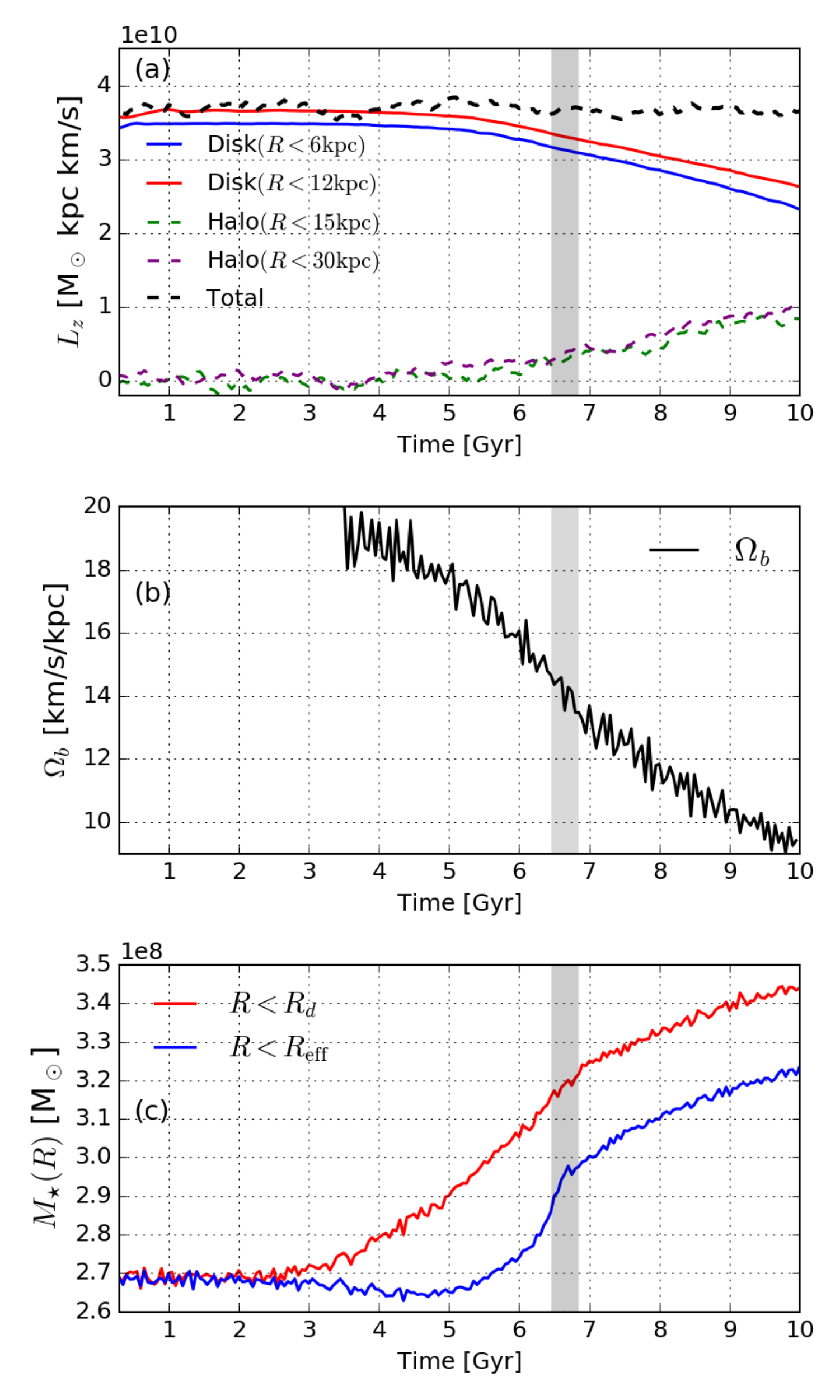}
\caption
{Evolution of (a) the angular momenta $L_z$ in the disk and the halo, (b) the bar pattern speed $\Omega_b$, and (c)
the enclosed disk masses $M_{*}$ within $1R_d$ and 1$R_{\rm eff}$  in model S1.
In (c), we subtract $2.5 \times 10^8\Msun$ from $M_{*}(R<R_{\rm eff})$ for a clear comparison. The shaded vertical strips  between $t=6.4$ and 6.9 Gyr mark the occurrence of the buckling instability. }\label{fig06dynamics}
\end{figure}

\subsection{Angular Momentum Transfer}

To explore dynamical friction between the bar and the halo, Figure \ref{fig06dynamics}(a) and (b) plot the time variations of the angular momenta $L_z$ in the disk and the halo, as well as the bar pattern speed $\Omega_b$. Soon after the bar forms, it begins to transfer angular momentum from disk to halo, keeping the total angular momentum almost constant. This is roughly consistent with the results reported by \cite{athanassoula03} who showed that the resonant interactions between the bar and the halo play an important role in the growth and slowdown rate of bars (see also \citealt{athanassoula05, athanassoula07, athanassoula14, sellwood16} and references therein). Note that the angular momentum exchange occurs continuously, regardless of the onset of the buckling instability, although its rate slows down as the bar becomes shortened. In fact, the presence of a \emph{live} halo is essential for the operation of the buckling instability that forms a peanut/boxy bulge \citep{berentzen06}.

As a consequence of the angular momentum transfer between the bar and the halo, the bar pattern speed gradually slows down from $\sim20$ to $\sim10\kms\kpc^{-1}$. A close inspection of Figure \ref{fig06dynamics}(b) reveals that the slowdown rate, $-d\Omega_b/dt$, increases with time until the onset of the buckling instability and becomes more or less constant afterwards. This is because the change in the pattern speed is closely related not only to the change in the angular momentum, but also to the change in the moment of inertia of the bar (e.g., \citealt{athanassoula03}). Indeed, the slowdown rate $-d\Omega_b/dt$ increases as the bar semi-major axis continuously increases from 1.5 to 2.0 kpc before the buckling instability. Since our standard model neither initially contains high rotational energy nor experiences a strong buckling instability, it does not exhibit a dramatic change in $d\Omega_b/dt$. We defer a more detailed discussion on this to Section \ref{s:ch4} where we present the cases with a stronger buckling instability (e.g., model H1).

As stellar particles lose angular momentum, they gradually migrate radially inwards to increase the enclosed mass $M_\star(R) = 2\pi \int_0^R \Sigma_\star R dR$ over time, as shown in Figure \ref{fig06dynamics}(c). Such bar-induced mass inflows were also observed in previous studies \citep{athanassoula02,athanassoula05, athanassoula16}. Similarly to $d\Omega_b/dt$, the mass inflow rate $-dM_\star/dt$ switches its slope abruptly at around the time of the buckling instability. Over 10 Gyr, the increase in the disk mass is $\sim27\%$ within $1R_d$: $\sim19\%$ and $\sim8\%$ before and after the buckling instability, respectively, suggesting that the mass inflow rate correlates well with the rate of the angular momentum transfer.

\subsection{Buckling Instabilities}\label{sec:buck}

In the preceding subsection, we have shown that our standard model is intrinsically unstable to forming a bar without environmental effects. It subsequently experiences buckling instability at $\sim6.4$--$6.9$ Gyr, which not only makes the bar smaller but also induces a transition in  $\omegadt$ and $\lzdt$. The bar becomes stronger again and regrows after the buckling instability is over. In this subsection, we use the velocity anisotropy  and the mean vertical velocity to analyze the first and ensuing buckling  instabilities.

\begin{figure}
\centering\includegraphics[angle=0,width=8.5cm]{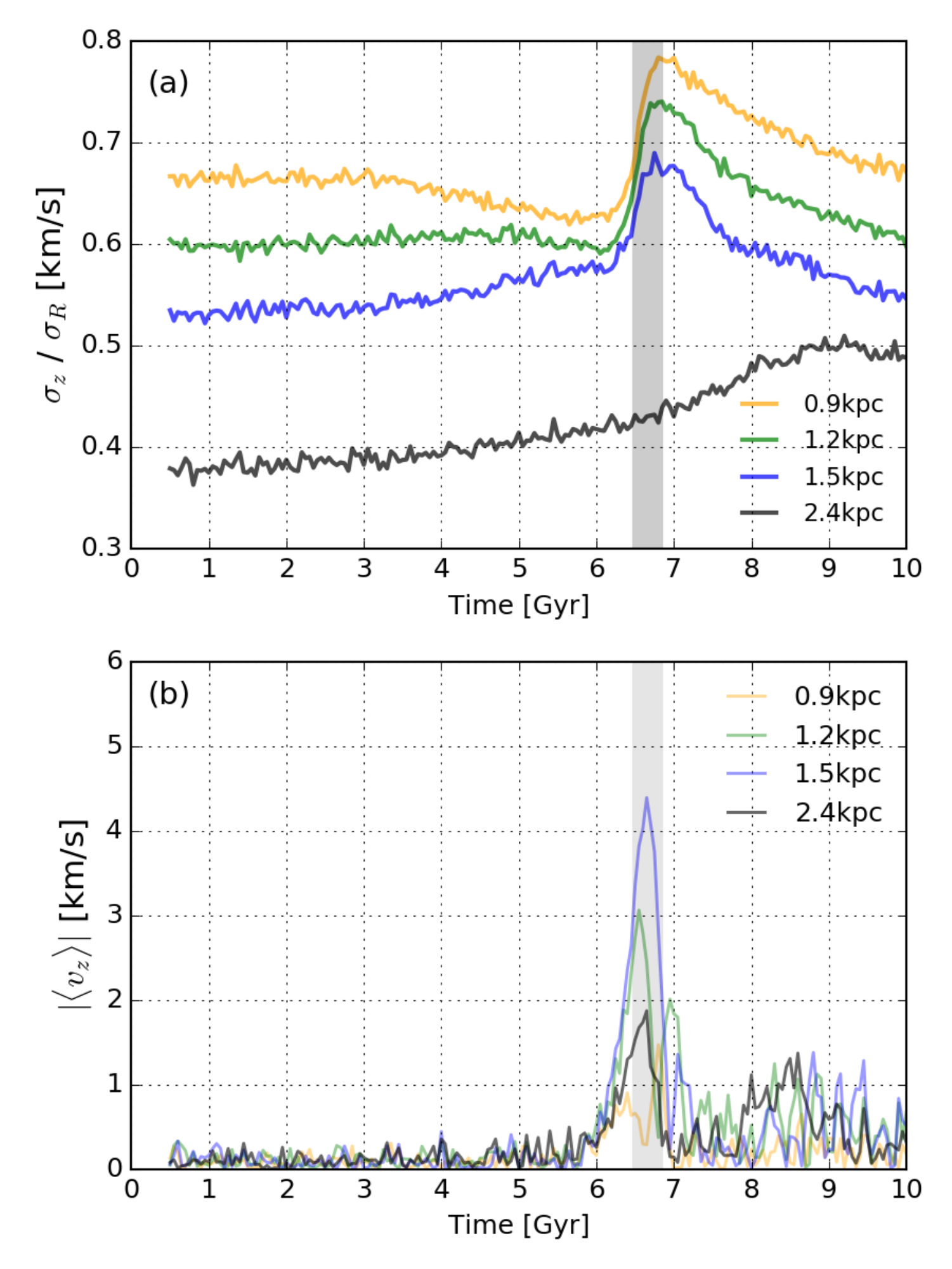}
\caption
{Evolution of the ratio of the velocity dispersions $\sigma_z/\sigma_R$ (top) and the absolute value of the azimuthally averaged vertical velocity $\absvz$ (bottom) as functions of time. The shaded vertical strips between $t=6.4$ and 6.9 Gyr mark the occurrence of the first buckling instability.}\label{fig07buckling}
\end{figure}

The local theory of buckling instability predicts that non-rotating, plane-parallel disks with $\sigma_z/\sigma_R \lesssim0.3$ are unstable to buckling out of the galactic plane, similarly to fire-hose instability (e.g., \citealt{toomre66,araki85}). For a realistic disk with radially varying $\sigma_z/\sigma_R$, the critical values are at $\sigma_z/\sigma_R\simeq 0.25-0.55$ in mid-disk regions \citep{raha91}, and at $\sim0.77$ in central regions \citep{sotnikova05}. \citet{martinez06} showed that the critical value in the central regions of their model disks is $\sim0.66$. These indicate that the critical value for buckling instability depends on the density and velocity structure of a disk-halo system.

Figure \ref{fig07buckling} plots the evolution of the velocity anisotropy  $\sigmaratio$ and the mean vertical velocity $\langle v_z \rangle$ of model S1.
The occurrence of the first buckling instability is apparent with a sudden increase of $\sigmaratio$ and $\absvz$ at $t\sim$6.4 Gyr. Similarly to Figure 6(a) of \cite{martinez06}, $\sigmaratio$ at $R=0.9\kpc$ in our model S1 decreases to exceed the critical value of $\approx0.63$, while its trend is different at outer radii: specifically, it stays almost constant at $R=1.2\kpc$ and increases slowly at $R=1.5\kpc$. The buckling instability is a global phenomenon that momentarily warps the entire bar into a U or V shape. These different temporal trends of $\sigmaratio$ indicate that the hypocenter of the buckling instability is located at small $R$ where the centrifugal force is very strong (e.g., \citealt{raha91}).

During the buckling instability (inside the shaded strip), $\sigmaratio$ at $R\lesssim1.5\kpc$ sharply increases: $\sigma_z$ is increased to thicken the disk and $\sigma_R$ is decreased to weaken the bar (see Fig.~\ref{fig05disp}).
In turn, the vertical heating stabilizes the system, and this is more effective at smaller $R$. After the buckling, $\sigmaratio$ declines again, approaching its initial value as $\sigma_R$ re-increases with the corresponding increase in the bar strength (Figs.~\ref{fig04barstrength} and \ref{fig05disp}).
On the other hand, $\sigmaratio$ at $R=2.4\kpc$  does not display a visible increase during the buckling instability. It continuously increases to $\approx 0.5$ at $t=9\Gyr$ with some fluctuations and then decreases slightly due to secondary buckling instabilities (see below).

The occurrences of buckling instabilities are more apparent in the temporal behavior of $\absvz$ shown in Figure \ref{fig07buckling}(b). A prominent peak centered at $t\sim6.65\Gyr$ corresponds to  the first buckling instability.
As a result of the bar warping, the outer regions near the edge of the bar ($R=1.2$ and $1.5\kpc$) visibly experience vertical buckling. The first buckling is able to drive bending oscillations even at $R=2.4\kpc$ where the bar is absent. Interestingly, multiple bucklings at all radii outside $R=0.9\kpc$ follow subsequently until the end of the run. These recurrent secondary bucklings are observed even at $R=1.2$ and $1.5\kpc$ because the first buckling instability was too mild to effectively dissolve the bar (see Section \ref{s:ch4} for further detail). A more vigorous first buckling would more vigorously thicken and shorten the stellar system, increasing $\sigma_{z}$ and making the inner regions stable to the recurrent buckling instabilities. (For instance, in model V1, which undergoes the first buckling instability more vigorously than model S1, the second buckling at $R=1.2\kpc$ is not observed in Figure \ref{fig09v2}(b).) At $R=2.4\kpc$, the secondary bucklings occur at $t=8$--$10\Gyr$, resulting in mild vertical thickening observed in the edge-on views at $t=8$ and $10\Gyr$ in Figure \ref{fig03image}.

To summarize,  in this section we construct a galaxy model that mimics VCC 856 representing a progenitor of the infalling dEdis, and evolve it for 10 Gyr.
Our results imply that such infalling dEdis, even without any external perturbation, are susceptible to bar formation. The vertical buckling instability is subsequently triggered to shorten the bar and to thicken the disk. The first buckling instability drives a dynamical transition in the pattern speed and mass inflow rate. Secondary buckling instabilities are observed in dwarf-sized disk galaxies even though their dark matter fraction is much smaller than that of normal disk galaxies. The secondary bucklings in our standard model lead to an X-shaped bulge, as in the Milky Way-sized model of \citet{martinez06}, although the X-shape is less prominent in our model due to a weaker bar regrowth after the first buckling.

\section{Effects of Varying Parameters}\label{s:ch4}

In this section, we conduct a comprehensive study on the effect of  galaxy parameters that vary within observational ranges. More specifically, we change the fraction of counter streaming stars, ratio of the velocity dispersions, disk scale height, halo concentration parameter, and the stellar mass.  We present how these parameters affect the stability of disks, and alter
the physical properties of bars that form and undergo the buckling instabilities.

\subsection{Counter Streaming Motion}\label{s:ch4v}

\subsubsection{Bar Formation and Evolution}

The presence of stars in counter streaming motions in real galaxies has been extensively reported in the literature \citep{bettoni90,merrifield94, bertola99, kannappan01a}, and available theories suggest that they can arise naturally due to gas infall and/or mergers during galaxy formation
\citep{thakar96, thakar98, algorry14}. In order to make the observed rotation curve consistent with the observed mass distribution (Fig.~\ref{fig01dynamics}(a)), we have to allow for the counter streaming motions in our models of dEdis. (see also \citealt{satoh80, yurin14}). To study the effect of $\vphi$ on the disk stability, we run the V models with differing $k$, while other parameters are the same as in model S1 (see Table \ref{tbl:models}).

\begin{figure}
\centering\includegraphics[angle=0,width=8.5cm]{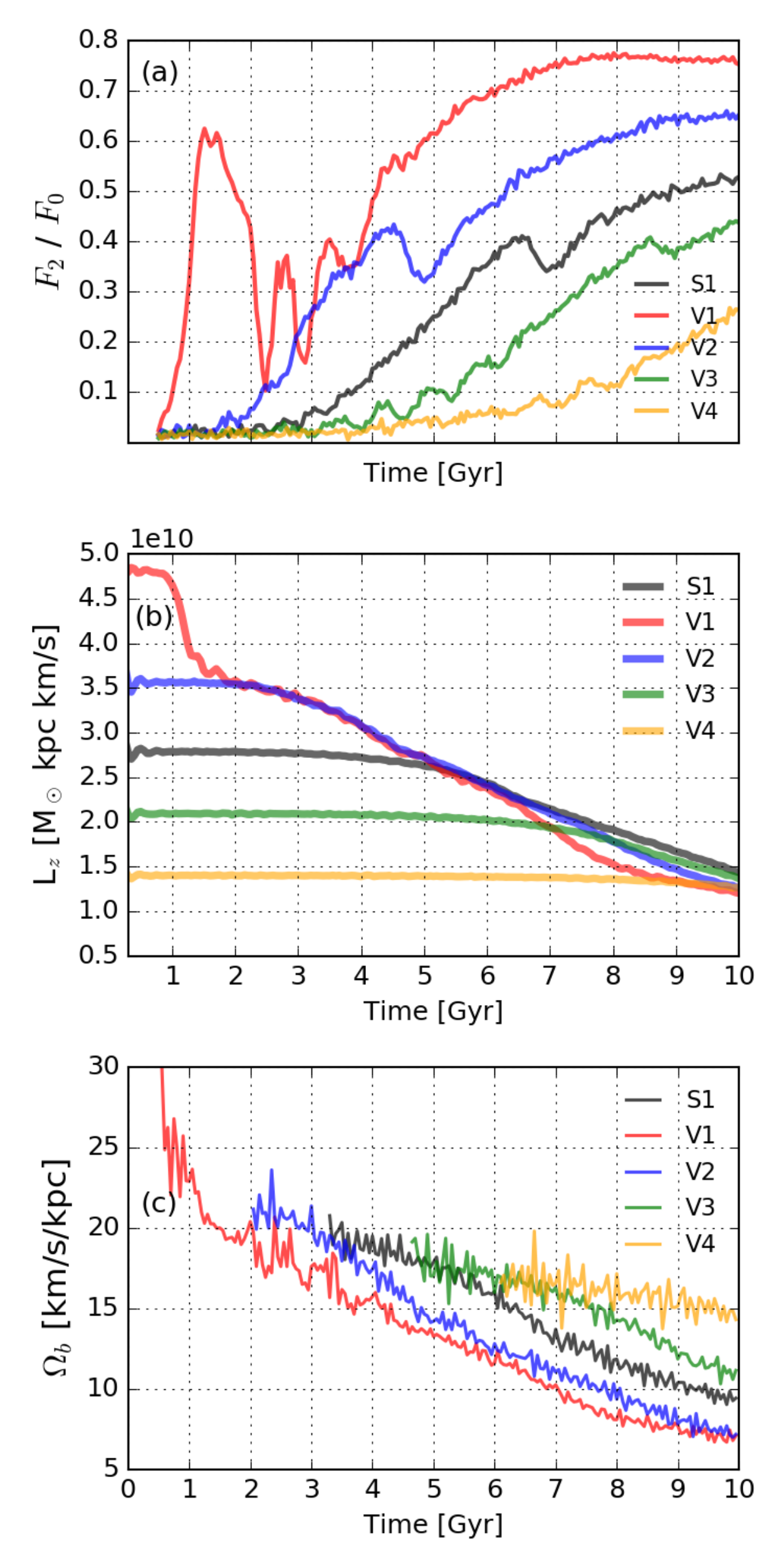}
\caption
{Evolution of (a) the bar strength $F_2/F_0$ at $R=1.5\kpc$, (b) the disk angular momentum $L_z$ within $R=3.1\kpc$, and (c) the bar pattern speed $\Omega_b$ in the V models as compared to model S1. }\label{fig08v1}
\end{figure}

\begin{figure}
\centering\includegraphics[angle=0,width=8.5cm]{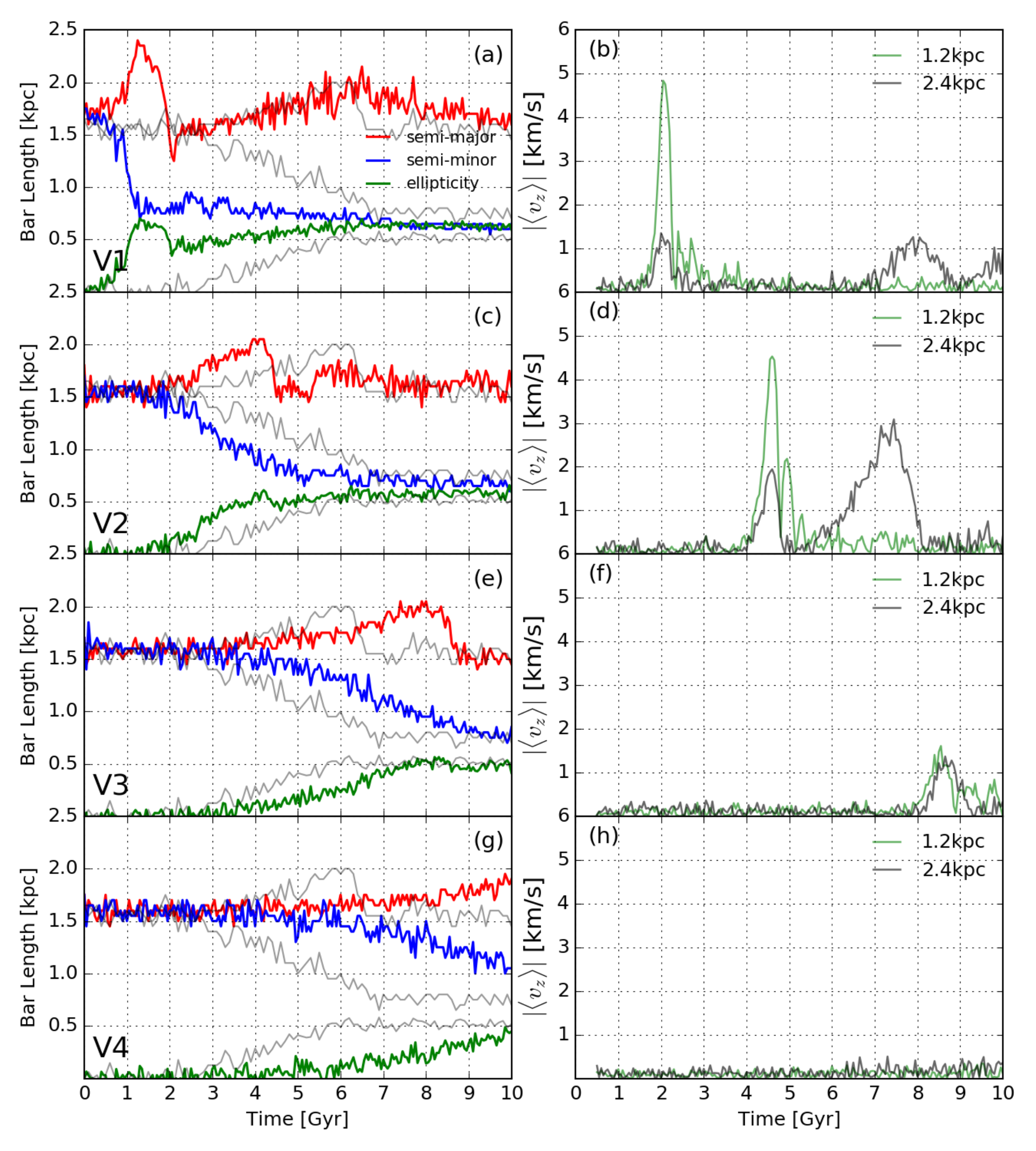}
\caption
{Evolution of the bar length (left) and the absolute value of the mean vertical velocity $\absvz$ (right) in V models. For comparison, the grey lines in the left panels draw the results of model S1.}\label{fig09v2}
\end{figure}

Figure \ref{fig08v1} plots the temporal variations of the bar strength $F_2/F_0$ at $R=1.5\kpc$,  the disk angular momentum $L_z$ inside $R=3.1\kpc$, and the bar pattern speed  $\Omega_b$ of V models in comparison with those in model S1. Clearly, a model with a higher $k$ is more unstable and more efficiently transfers the angular momentum to the halo, forming a stronger bar at an earlier time. Except for model V4,  all V models experience a buckling instability in 10 Gyr, and the buckling is more violent in a more unstable disk. The temporal decrease of the bar pattern speed  is caused by the angular momentum transfer, consistent with the linear theory of \citet{athanassoula03}.

\begin{figure*}
\includegraphics[angle=0,width=17cm]{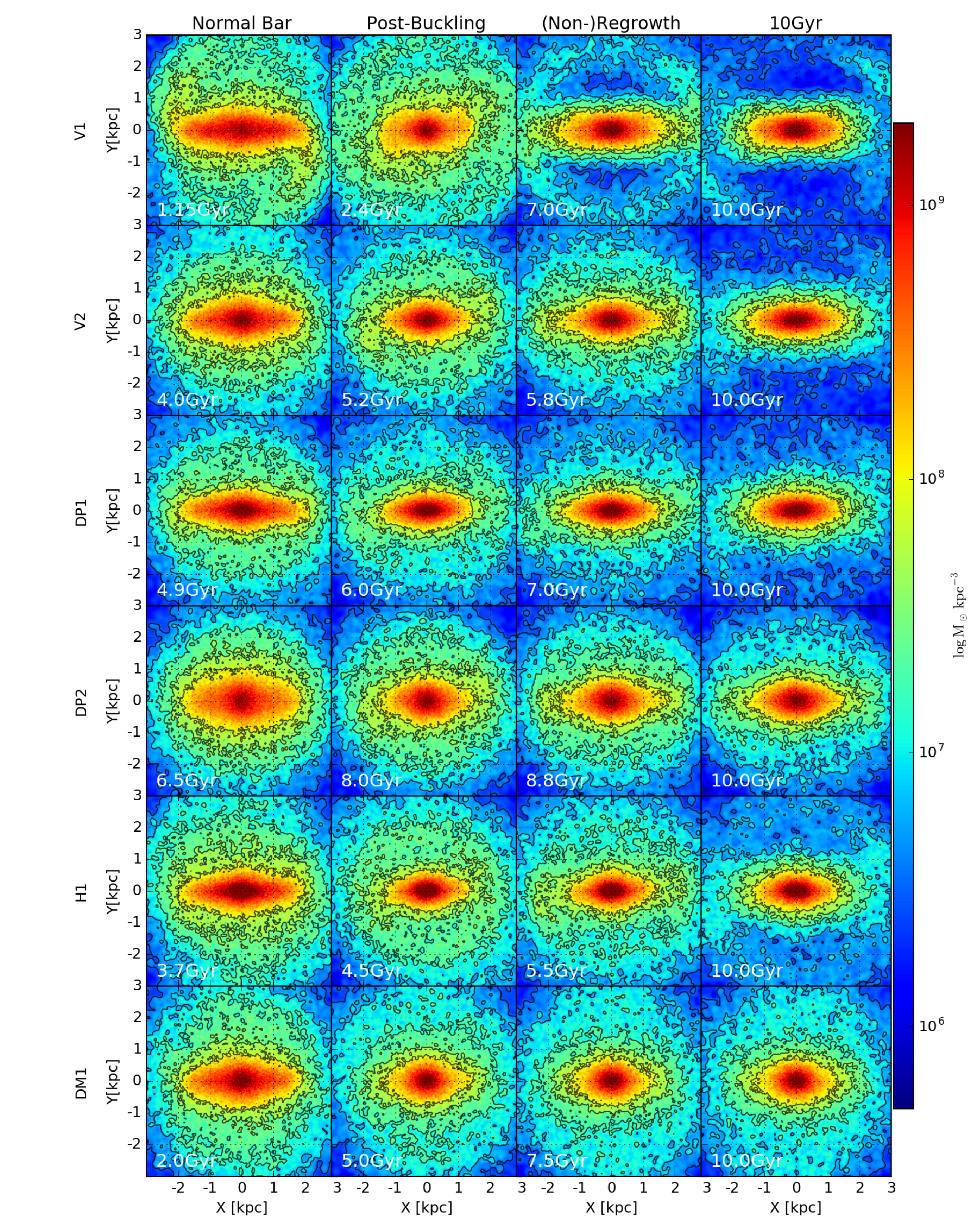}
\caption
{Density snapshots at the $z=0$ plane of the models that experience the buckling instability at some selected epochs corresponding to normal, post-buckling, (non)-regrowth, and 10 Gyr from left to right panels. The specific time of each panel is given at the lower left corner. Except for model DM1, all other models experience the second buckling instability.
}\label{fig10face}
\end{figure*}

\begin{figure*}
\includegraphics[angle=0,width=17cm]{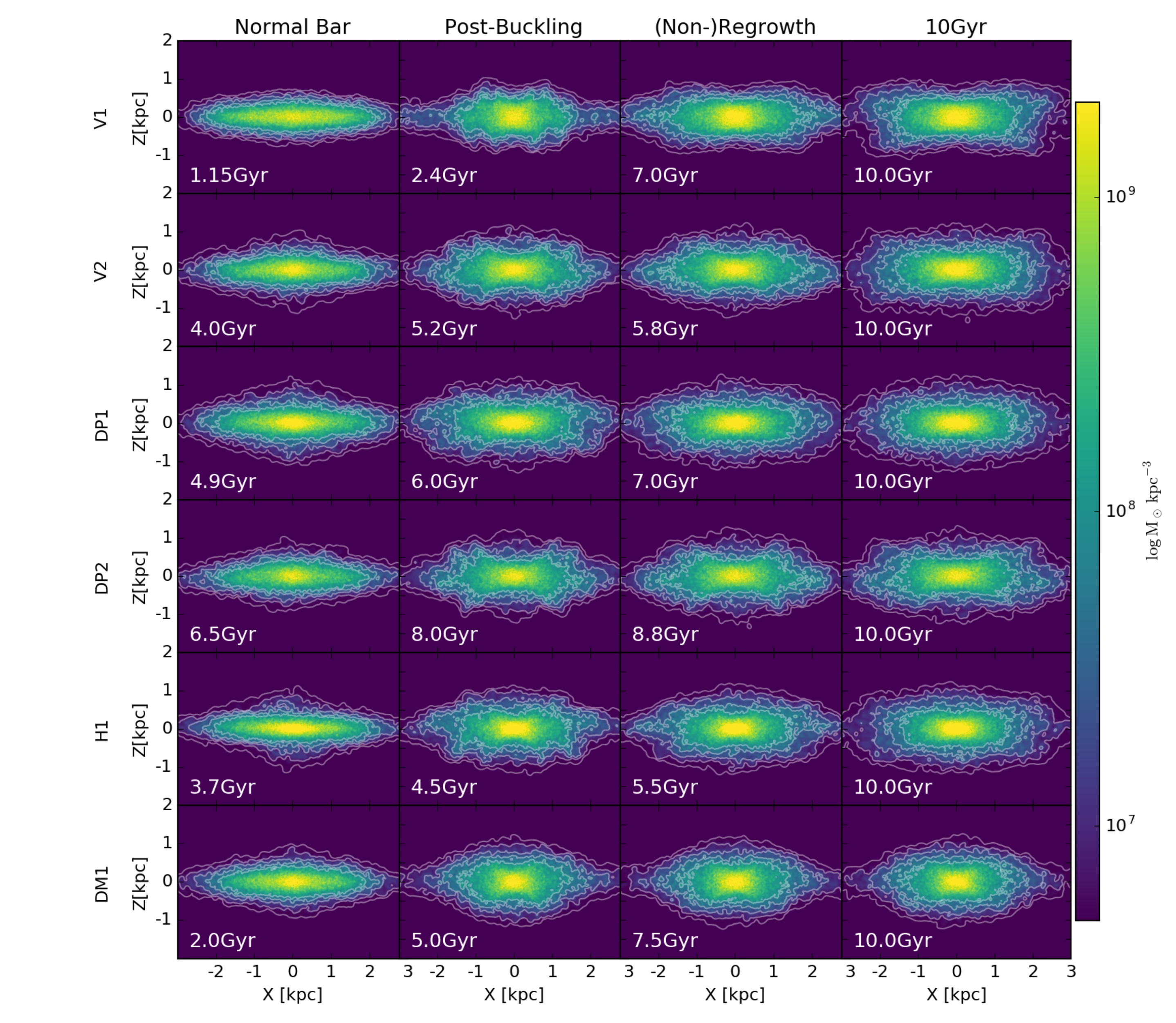}
\caption
{Same as Figure \ref{fig10face} but for the density snapshots at the $y=0$ plane.
}\label{fig11side}
\end{figure*}

In terms of $Q_T$, models V1 -- V4 should have the same stability character. But, our numerical results show that this is not the case.
There is another stability criterion, $Q_E \equiv v_{\rm max}/\sqrt{GM_d/R_d} > 1.1$, with $v_{\rm max}$ being the maximum rotation velocity
proposed by \citet{fall80} and \citet{efstathiou82}, based on  $N$-body simulations of two-dimensional disks under a rigid halo. The $Q_E$ values of our V1 to V4 models are 0.62, 0.48, 0.28, and 0.18, respectively, suggesting that model V4 should be the most unstable among them. This again contradicts our numerical results, in that model V1 is the most unstable. \cite{athanassoula08} already noted that the $Q_{E}$ criterion is too simplistic to provide the actual stability of disks.

The order of the bar formation time in V models appears to be inversely proportional to the $L_z$ of the initial disk (Fig.~\ref{fig08v1} (b)).
Model V1 forms a bar first, and its strength exceeds $\sim0.6$ at $1.2\Gyr$, implying that disks with higher $L_z$ tend to be more unstable and form a stronger bar. Model V2, which contains less counter streaming motion than our standard model, forms a prominent bar earlier at $4.0\Gyr$. As discussed in Section \ref{s:ch3}, the first episode of buckling instability induces a dynamical transition in $L_z$ by altering $\lzdt$ that closely correlates with $\omegadt$. However, in models V1 and V2, even after the first buckling instability that is stronger than in model S1, $-\lzdt$ does not decrease considerably, eventually aiding the bar regrowth in models V1 and V2.

Figure \ref{fig09v2} plots the evolution of the bar length (left) and the
mean vertical velocity $\vz$ at $R=1.2$ and $2.4\kpc$ (right) of the V models.
The bars in models V2 -- V4 are no longer than $\sim2.1\kpc$. In contrast, model V1 forms a long bar and undergoes a vigorous buckling instability at $\sim2.0\Gyr$, significantly shortening the bar from $\sim2.4$ to $\sim1.5\kpc$. After the buckling, the bar slowly regrows up to $7\Gyr$ and then decreases again. Note that the slow bar regrowth in model V1 induces a mild acceleration in the decay of $\Omega_b$ and $L_z$ before the second buckling instability, which shortens the bar again (Fig.~\ref{fig08v1}(b) and (c)).

The temporal behavior of $\vz$ is useful for determining the epochs of buckling instabilities. Considering the height/width of the peaks in $\vz$ as well as the corresponding decrease in the bar length, we find that a more unstable disk undergoes more vigorous buckling instability. In model V1, the first peak of $\absvz$ occurring at $t\sim2\Gyr$ corresponds to a sharp decrease in the bar strength and length. Between $t=7$ and $9\Gyr$, the second buckling instability is observed at $R=2.4\kpc$, accounting for the second episode of the bar shortening after $\sim7\Gyr$. Interestingly, model V1 suffers a third buckling instability at $R=2.4\kpc$. Model V2 also displays two peaks in $\absvz$, while model V3 experiences a weak buckling instability only once.
Model V4 forms a bar, but it is sufficiently stable not to experience a buckling instability until the end of evolution.

Similarly to model V1, model V2 also experiences the second buckling instability, but earlier and more prominently than model V1. This is because the first buckling in model V2 is weaker and thus causes less damage to the bar, providing a more favorable environment for the regrowth. The bar regrowth in model V2 thus follows rapidly at $t\sim5.5\Gyr$ and triggers the second buckling, but only in the outer regions ($R\sim2.4\kpc$) that are unaffected by the first buckling. Although \cite{martinez06} demonstrated that the regrowth is aided by the angular momentum exchange between the bar and the halo, our results suggest that the effect of the first buckling is also an important factor for the regrowth and recurrent buckling instability.

Figures  \ref{fig10face} and \ref{fig11side} plot the face-on and side-on snapshots of the disk density at the $z=0$ and $y=0$ planes, respectively, for some selected epochs corresponding to normal bar, post-buckling, regrowth or non-regrowth and 10 Gyr from left to right. The first two rows are for models V1 and V2. A close inspection reveals that the bar length indeed becomes slightly longer after the first buckling, but the second buckling shortens the bar again. The edge-on views demonstrate that the disks are thickened vertically and take on a peanut/boxy shape after the first buckling. The secondary bucking makes the disk thicker in the outer regions around $R=2.4\kpc$, which is more pronounced in model V2, which has a higher $\absvz$. As a result, an X-shaped bulge is more evident in model V2. This is consistent with \cite{martinez06} who showed that the first buckling leads to a peanut/boxy shape while the second buckling yields an asymmetric shape in relatively outer regions, eventually leading to an X-shaped bulge (see Fig.3 of \citealt{martinez06}). As the third buckling with a small $\absvz$ subsequently follows at $t~9.5\Gyr$ in model V1, its edge-on view still displays such a faint asymmetry at $t=10$ Gyr.

\begin{figure}
\centering\includegraphics[angle=0,width=8.5cm]{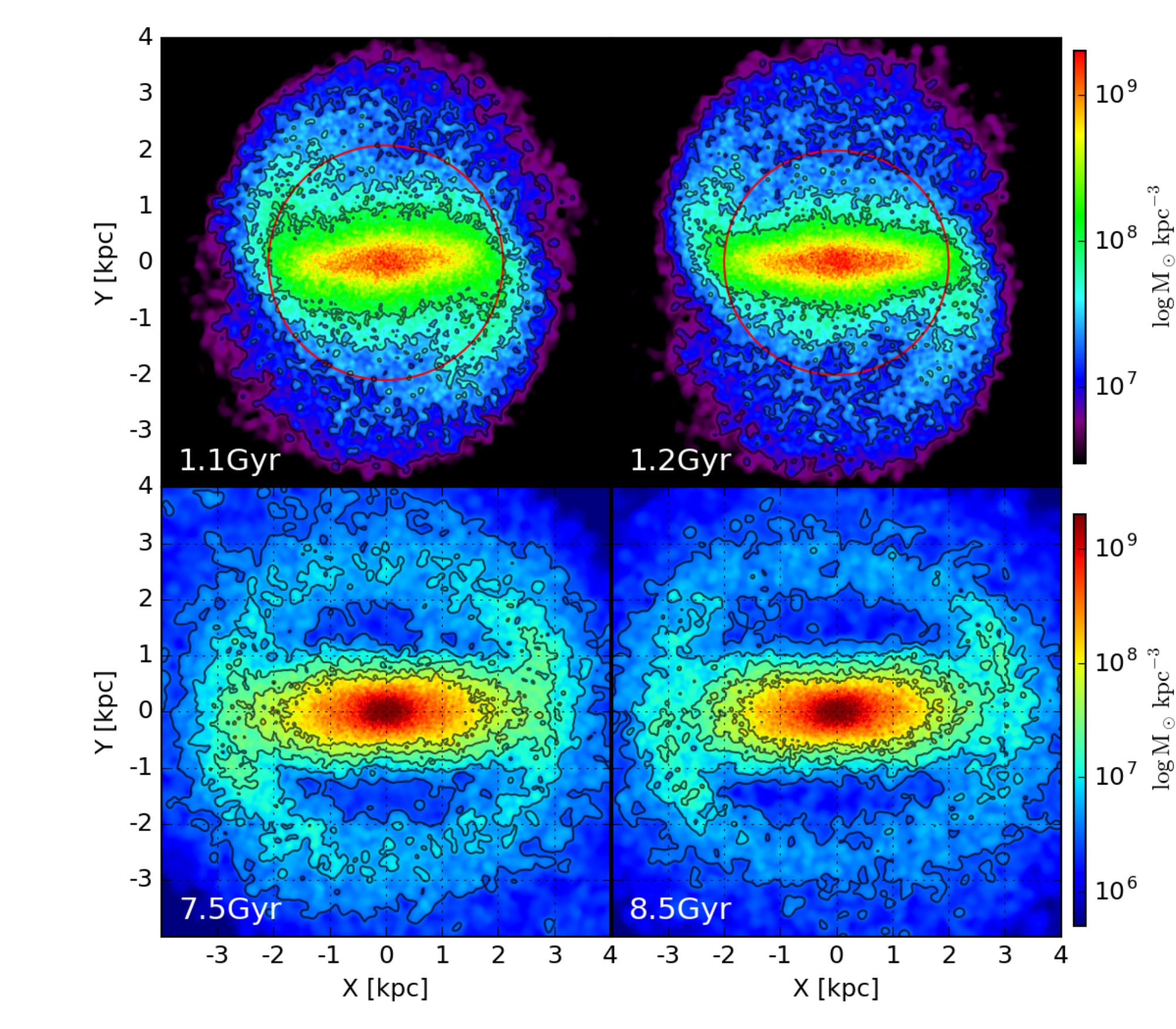}
\caption{Face-on views of model V1 at an epoch time when the barred-spirals or an inner stellar ring are observed. The circles in the upper panels mark the corotation radius. The bar rotates in the counterclockwise direction.}\label{fig12ring}
\end{figure}

\subsubsection{Manifold-driven Spirals and Inner Ring}

In addition to the bar, model V1 also possesses spiral arms extending from the bar ends at $t=1.15\Gyr$ and an inner stellar ring encompassing the bar at $t=7.0\Gyr$ (Fig.~\ref{fig10face}). The proposed formation mechanism of barred spiral arms is the ejection of stars in the vicinity of bar ends into chaotic motions when two Lagrangian points $L_1$ and $L_2$ become unstable  \citep{romero06,romero07,athanassoula09a,athanassoula09b,athanassoula10}.
These manifold-driven stellar trajectories also account for the formation of stellar rings in various shapes, depending on the bar properties. Manifold-driven spirals in a realistic $N$-body simulation were first observed by \cite{athanassoula12}, although their duration was too short to maintain the spirals and ring shape of longer than 1 Gyr \citep{athanassoula12, lokas16a}.

To more vividly display the spirals and inner ring, Figure \ref{fig12ring} plots face-on views of model V1 at $t=1.1, 1.2, 7.5,$ and 8.5 Gyr. The red solid circle in the first row draws the corotation radius (CR). The spirals begin to appear from $t=1.1\Gyr$ when the bar length reaches the CR. The face-on view at $t=1.2\Gyr$ more prominently displays the barred spirals as the bar length and strength attain their maximum. Thus, stellar particles in the bar outside the CR are ejected from the bar, becoming spirals. This in turns results in the decrease in the bar strength and length until $t\sim1.8\Gyr$, shortly before the onset of the first buckling instability, which is consistent with \cite{lokas16a}, who showed that the manifold-driven spirals persist only in between the bar formation and the buckling instability. We find no evidence of spirals in the other models. The primary reason may be that their bars always reside within the CR, indicating that a bar should be sufficiently long to form manifold-driven spirals.

Figure \ref{fig12ring} shows that model V1 has an inner ring emanating from the ends of the bar. It is more or less circular at $t=7.5$ Gyr and becomes elongated at a late time. The inner ring survives until the end of the run, although some parts of the ring fade gradually. This $\Theta$-shaped morphology closely resembles model MH of \cite{athanassoula02} who demonstrated that the ring formation depends on the mass ratio between the disk and the halo.  The fact that the inner ring is produced only in model V suggests that its formation requires a very strong bar, perhaps with $F_2/F_0 > 0.7$. Similarly to spirals, the inner stellar ring can be a structure guided by the manifolds associated with the unstable Lagrangian points located near the bar ends \citep{athanassoula09b, athanassoula10}.

\subsection{Ratio of Velocity Dispersions}\label{s:ch4dp}

\begin{figure}
\centering\includegraphics[angle=0,width=8.5cm]{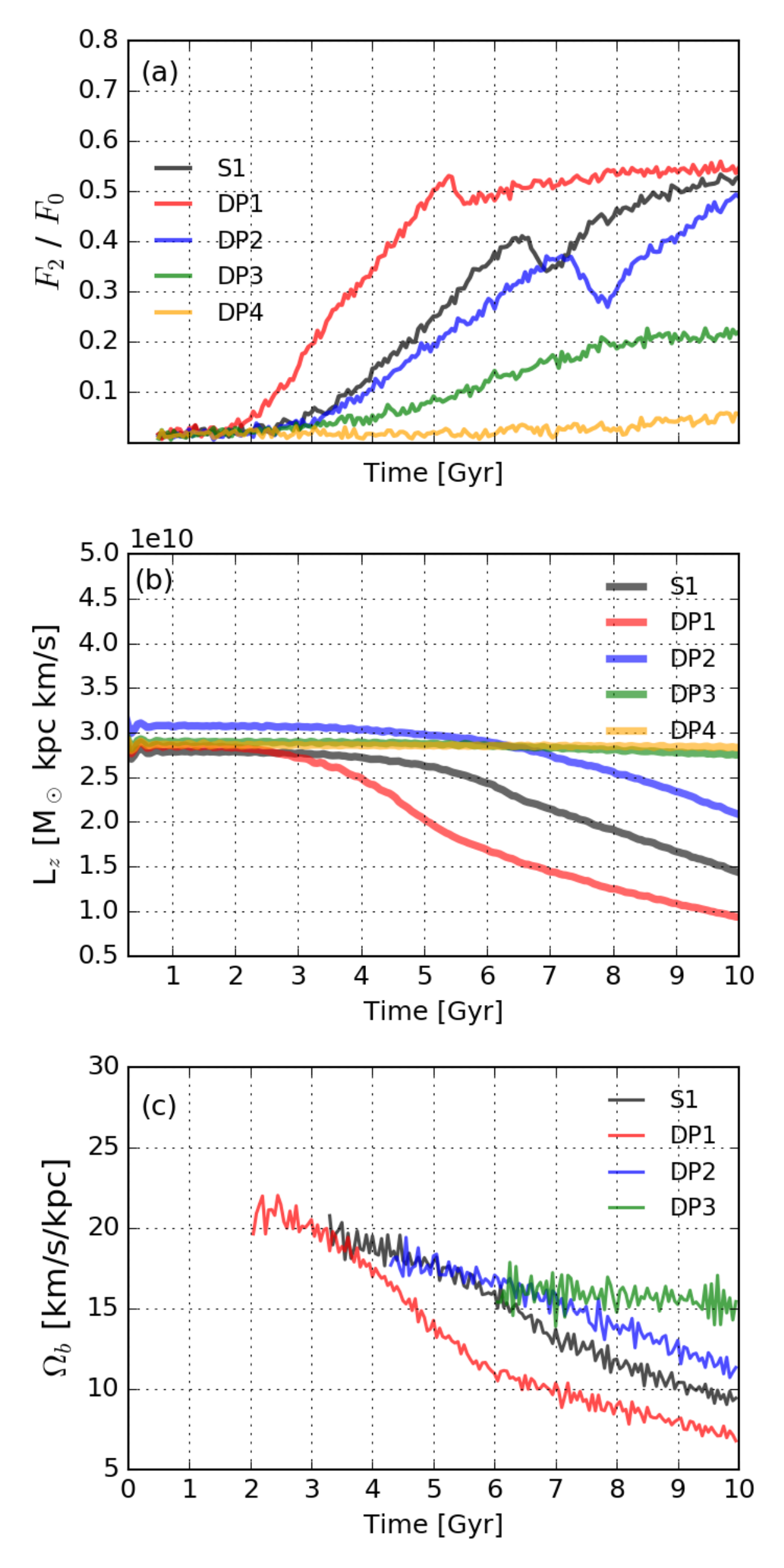}
\caption
{Evolution of (a) the bar strength $F_2/F_0$ at $R=1.5\kpc$, (b) the disk angular momentum $L_z$ within $R=3.1\kpc$, and (c) the bar pattern speed $\Omega_b$ in the DP models as compared to model S1.}\label{fig13dp1}
\end{figure}

\begin{figure}
\centering\includegraphics[angle=0,width=8.5cm]{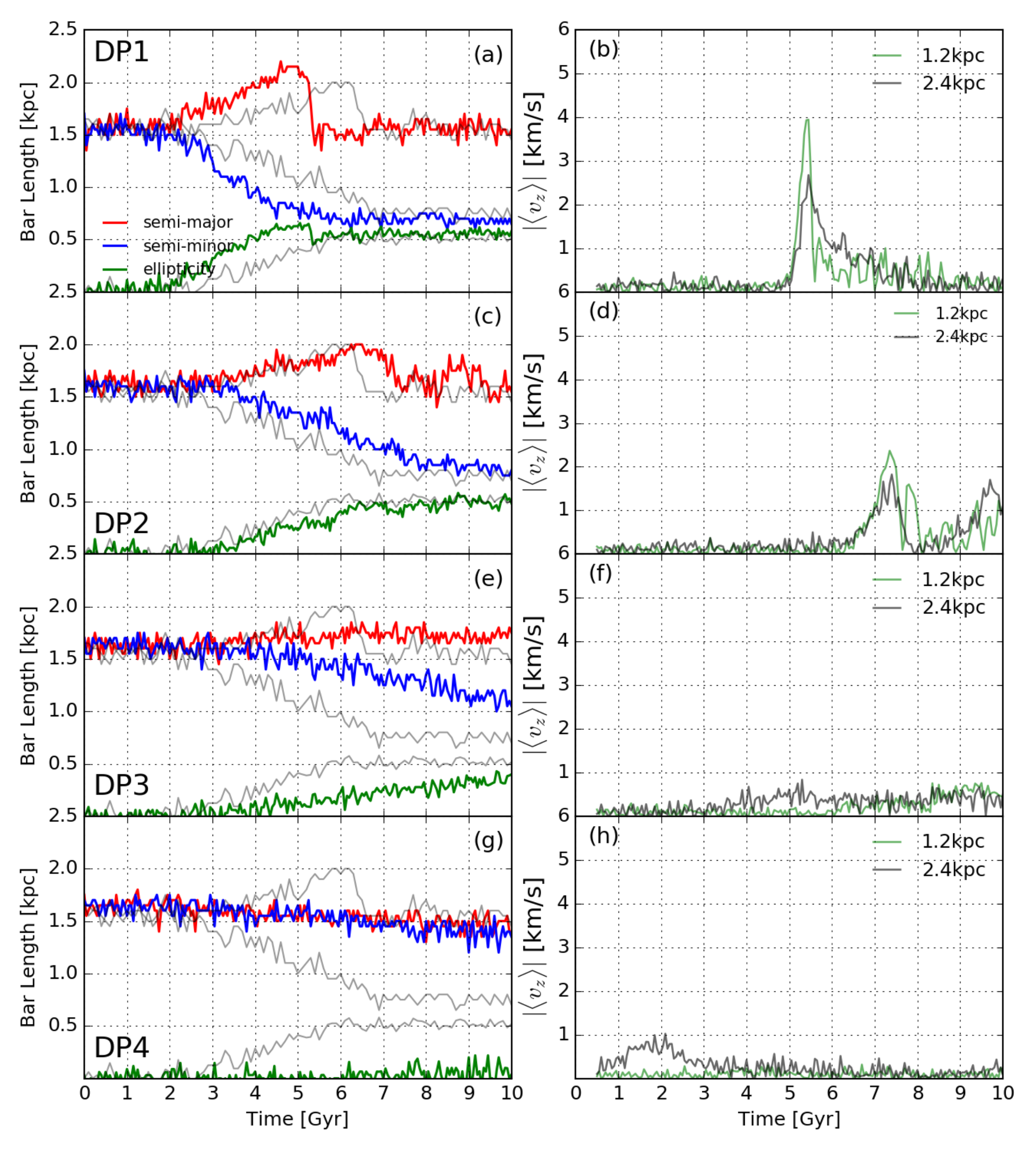}
\caption
{Evolution of the bar length (left) and the absolute value of the mean vertical velocity $\absvz$ (right) in the DP models.  For comparison, the grey lines in the left panels draw the results of model S1.}\label{fig14dp2}
\end{figure}

The ratio of the velocity dispersions is an important parameter, not only in the disk stability, but also in the buckling instability \citep{araki85, raha91}.
The working value for our standard model is $\sigma_{R}/\sigma_{z}=1.25$, which is adopted from \cite{lisker09} following the method of \cite{gerssen00}.
To examine its effect on the bar formation and buckling instability, we vary $f_{R}$ in Equation (\ref{eq:disp}) to construct four DP models with different $\sigma_{R}/\sigma_{z}$ (see Table \ref{tbl:models}). For fixed $\sigma_{z}$, increasing $f_{R}$ increases $\sigma_{R}$ and thus $Q_T$.

Figure \ref{fig13dp1} plots the temporal variations of $F_2/F_0$, $L_z$, and $\Omega_b$ of the DP models in comparison with model S1,
while Figure \ref{fig14dp2} plots the evolution of the bar length and $\vz$ of the DP models. As expected, a model with smaller $\sigma_{R}/\sigma_{z}$ forms a stronger bar earlier, and thus model DP1 with $\sigma_{R}/\sigma_{z}=1.0$ forms the thinnest and longest bar. As can be see in the leftmost panels of Figure \ref{fig10face}, models with larger $f_R$ have a shorter and more rounded bar. This is consistent with \citet{athanassoula86} who suggested that larger random motions can make a bar wider. While the bars in models DP1, S1, and DP2 with $\sigma_{R}/\sigma_{z} \lesssim 1.5$ become temporarily weaker due to the first buckling instability, the bar strength in model DP3 with $\sigma_{R}/\sigma_{z}=1.75$ stays almost constant at $\gtrsim 8$ Gyr. The bar semi-major axis in model DP3 does not increase considerably, while its semi-minor axis decreases slowly. In model DP4 with $\sigma_{R}/\sigma_{z}=2.0$, the bar instability is suppressed due to random motions, albeit not completely. Although all the DP models begin from nearly identical initial $L_z$, it is evident that colder disks lose a larger amount of angular momentum, resulting in a higher $-\omegadt$ before the buckling instability. After the buckling instability, $-\omegadt$ becomes nearly constant in all DP models that form a bar.

The presence of the buckling instabilities can be easily recognized in the temporal changes of $\absvz$ shown in Figure \ref{fig14dp2}. We find that the critical value $\sigma_{z}/\sigma_{R}\approx0.63$ at $R=0.9\kpc$ for the buckling instability is identical in all the models with the same density structures (S1, V, and DP models), although the buckling instability occurs later in hotter disks due to the inefficient angular momentum transfer \citep{athanassoula03}. Also, while all the DP models have nearly the same rotational kinetic energy, the buckling instability is more vigorous in colder disks. In models DP1 and DP2, the buckling instability is sufficient to considerably dissolve a bar. In model DP3, mild buckling instabilities are observed soon after the bar formation, which is too weak to stabilize the disk that undergoes repeated bucklings until the end of the run. Model DP4, with small $\sigma_z/\sigma_R$, displays a slight warp in the outer disk, without forming a noticeable bar.

\subsection{Disk Thickness}\label{s:ch4h}

\begin{figure}
\centering\includegraphics[angle=0,width=8.5cm]{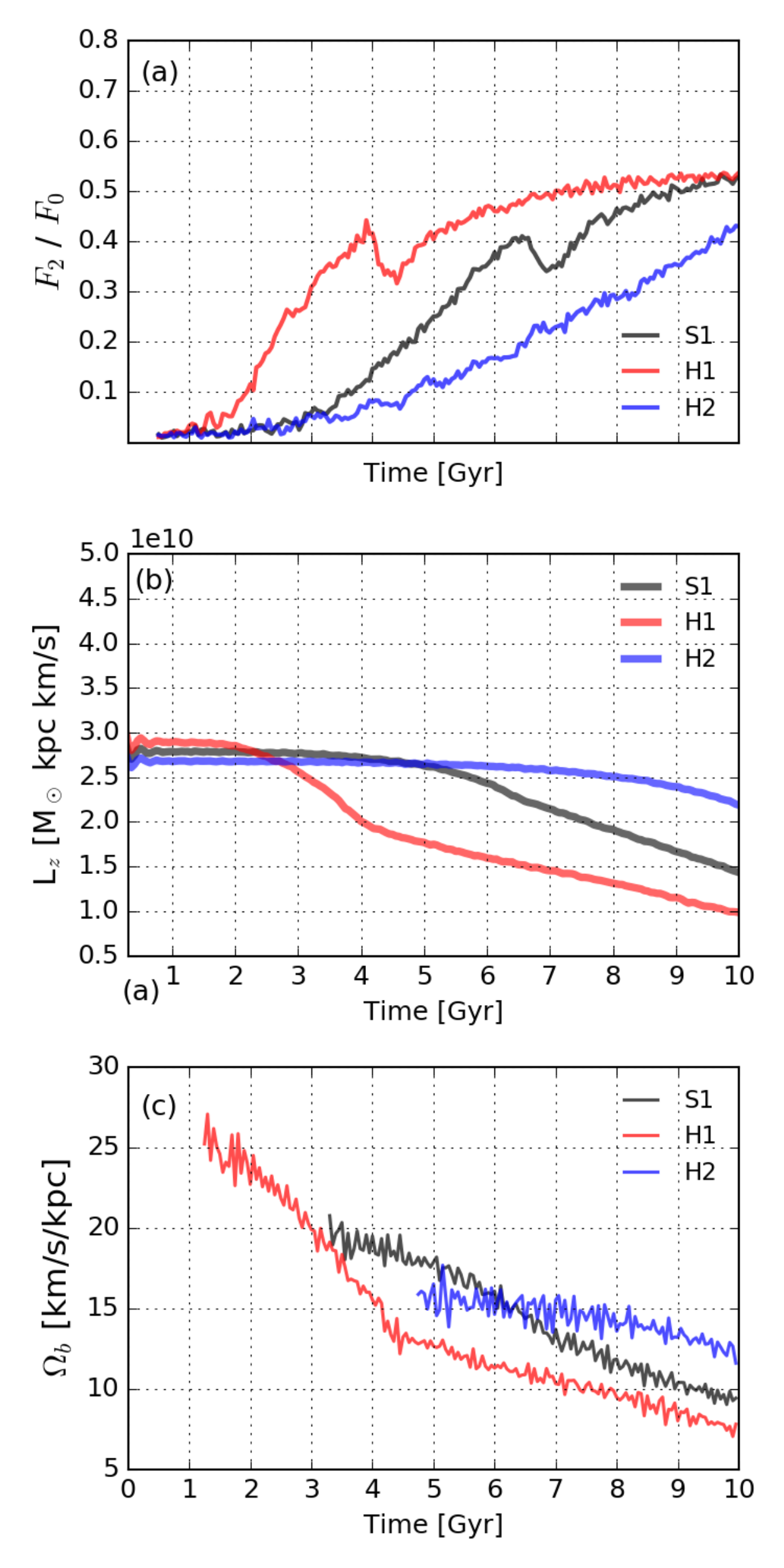}
\caption
{Evolution of (a) the bar strength $F_2/F_0$ at $R=1.5\kpc$, (b) the disk angular momentum $L_z$ within $R=3.1\kpc$, and (c) the bar pattern speed $\Omega_b$ in H models as compared to model S1.}\label{fig15h1}
\end{figure}

\begin{figure}
\centering\includegraphics[angle=0,width=8.5cm]{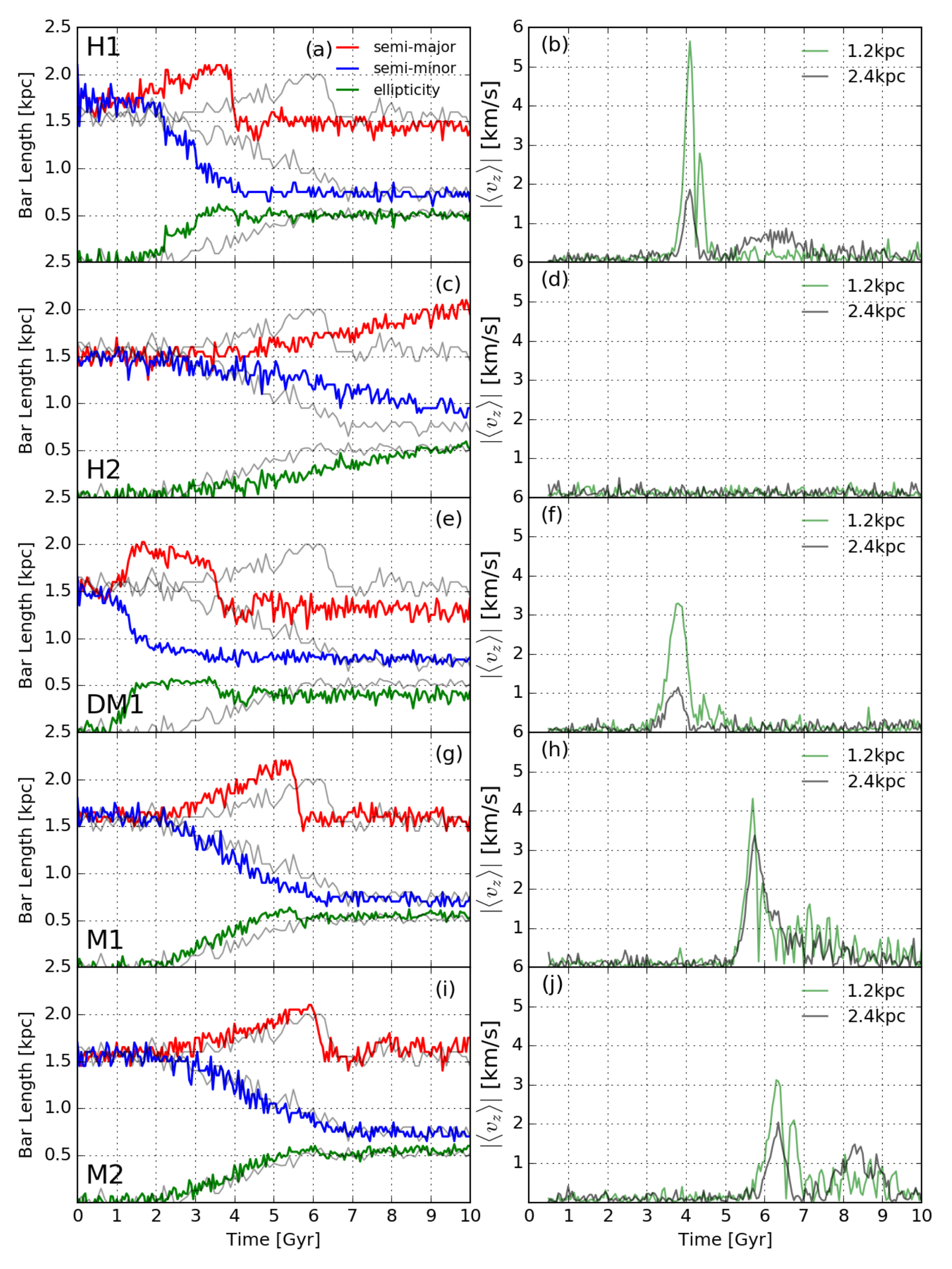}
\caption{Evolution of the bar length (left) and the absolute value of the mean vertical velocity $\absvz$ (right) in H, DM, and M models. For comparison, the grey lines in the left panels draw the results of model S1.}\label{fig16h2}
\end{figure}

The scale height of our standard model ($z_{d}/R_{d}$=0.33) is adopted from \cite{lisker09}. As the error range in the scale height of VCC856 is not known observationally, we simply vary the axial ratio by 0.1 from model S1.  With the other parameters fixed, we examine the sole effect of $z_d/R_d$ on bar instability and the vigor of buckling instability.

Figure \ref{fig15h1} plots the temporal variations of $F_2/F_0$, $L_z$, and $\Omega_b$ of H models in comparison with model S1.
The top two rows of Figure \ref{fig16h2}(a)--(d) plot the evolution of the bar length and $\vz$ in H models. Model H1 with $z_d/R_d=0.23$ is more unstable to forming a bar earlier than model H2 with $z_d/R_d=0.43$. It undergoes the first buckling instability at $t\sim4\Gyr$, which  decreases the bar length from $\sim2.2\kpc$ to $\sim1.5\kpc$ and a faint secondary instability with $\absvz<1.0$ km s$^{-1}$ at $R\sim2.4\kpc$ near $t=5$--$6 \Gyr$, without any sign of regrowth afterwards. The second buckling in model H1 is less visible than that in model S1 since its first buckling instability is more prominent, as discussed in Section \ref{s:ch4v}. Therefore, a bar in a thinner disk undergoes a more vigorous buckling, which considerably shortens the bar length and provides a hostile condition for the bar regrowth. On the other hand, model H2 does not undergo a buckling instability until the end of the run.

The edge-on views of model H1 are shown in the second row from the bottom of Figure \ref{fig11side}. As discussed in Section \ref{s:ch4v} and \cite{martinez06}, the first buckling instability leads to a peanut/boxy shape at $t=5.5\Gyr$. However, the vertical thickening around $R\sim2.4\kpc$ triggered by the second buckling is not as effective as that in model V2, forming a weak X-shaped bulge.

Compared to model S1, the bar pattern speed in model H1 is initially higher, and decays more rapidly due to higher $-\lzdt$. However, after the buckling instability, $-\omegadt$ in model H1 becomes almost constant at a slightly smaller value than that of model S1. Since the fraction of the counter streaming motions is the same in the S1 and H models, we attribute this difference to the shorter bar length in model H1 induced by a more vigorous buckling instability.

\subsection{Dark Matter Fraction}\label{s:ch4dm}

\begin{figure}
\centering\includegraphics[angle=0,width=8.5cm]{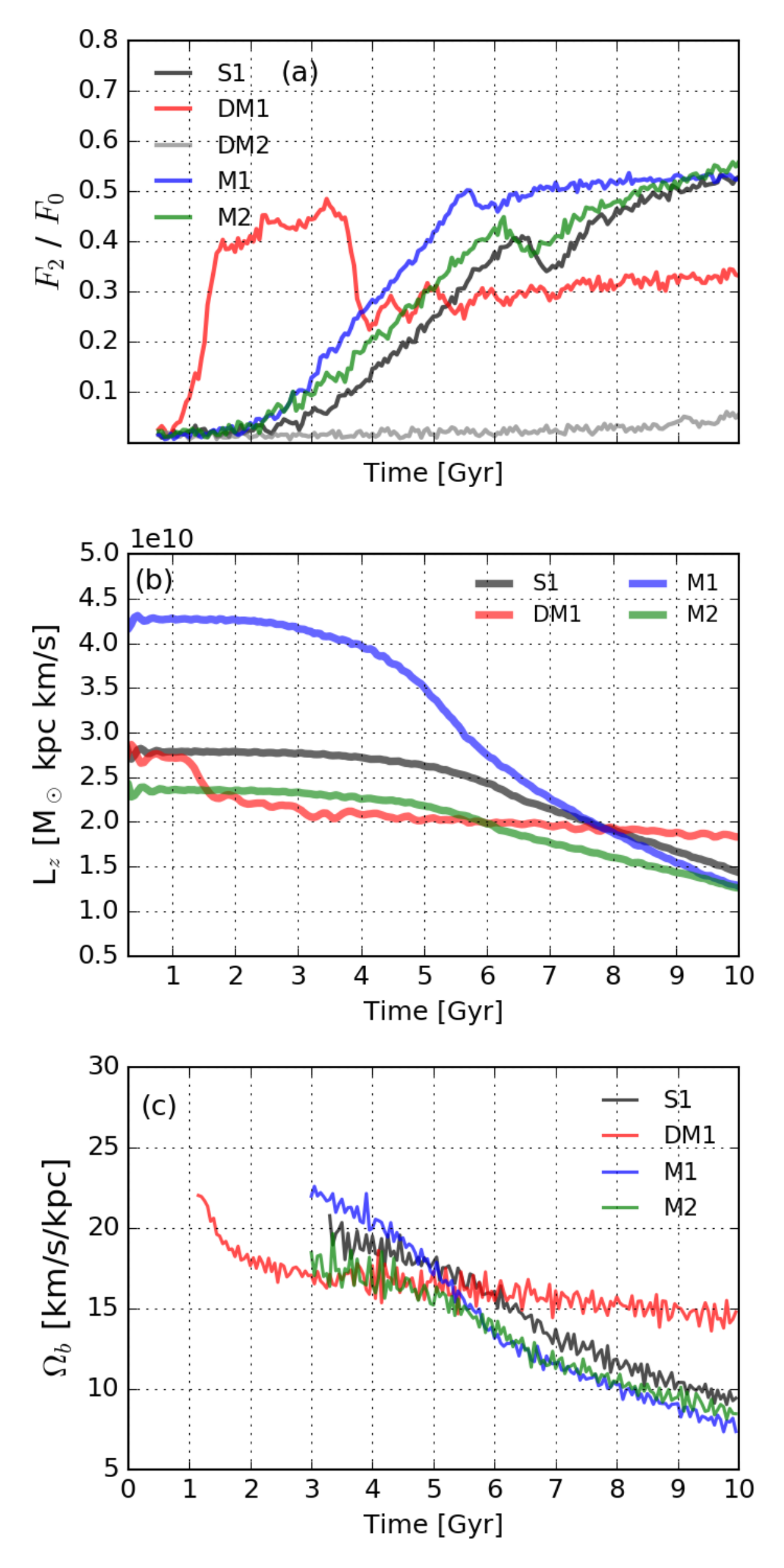}
\caption
{Evolution of (a) the bar strength $F_2/F_0$ at $R=1.5\kpc$, (b) the disk angular momentum $L_z$ within $R=3.1\kpc$, and (c) the bar pattern speed $\Omega_b$ in DM and M models as compared to model S1.
}\label{fig17dmm}
\end{figure}

To explore the effects of $f_\text{DM}$ on the disk stability, we vary the halo concentration in Equation \eqref{eq:cc}, while changing the $k$ parameter to obtain similar $\vphi$. Model DM1 with a lower halo concentration contains less counter streaming motion ($c=2$ and $k=0.8$) than model DM2, which has a more compact halo and more counter streaming particles ($c=20$ and $k=0.55$). Our choice of $c$ is to satisfy the error range of VCC856, $f_\text{DM}=0.33 \pm 0.26\%$ within $R_\text{eff}$ \citep{toloba14}.

The temporal variations of  $F_2/F_0$, $L_z$, and $\Omega_b$ of DM models are plotted in Figure \ref{fig17dmm}. It is well known that the presence of a compact halo stabilizes a disk against self-gravitating perturbations by breaking the feedback loop of a swing amplifier \citep{toomre81}. Indeed, Figure \ref{fig17dmm}(a) shows that model DM1, with a less compact halo, rapidly forms a prominent bar at $t=1.5\Gyr$, while model DM2, with a concentrated halo, remains stable until the end of the run. Model DM1 experiences buckling at $t\sim3.5\Gyr$, resulting in a significantly shortened and weakened bar. After the buckling, the bar immediately becomes boxy/peanut-shaped (Fig.~\ref{fig10face}). The bar strength subsequently increases with time, but only slightly as compared to model S1, without much change in the bar size and shape. From the temporal behavior of $\vz$  shown in Figure \ref{fig16h2}(f), the recurrence of the buckling instability is not evident in model DM1.

Figure \ref{fig17dmm}(b) shows that model DM1 lacks in the angular momentum transfer due primarily to the insufficient interactions between the bar and the halo \citep{athanassoula03, berentzen06}, simply caused by insufficient dark matter around the disk. Once it rapidly forms a prominent bar with $F_2/F_0\sim0.4$ at $t\sim1.5\Gyr$, its angular momentum loss rate becomes very small, as is the slowdown rate of the bar. Such deficient bar-halo interaction also yields a low growth rate in the bar strength. This causes a transitional phase ($t=1.5$ -- 3.5 Gyr) in model DM1 during which the bar strength slowly increases before the onset of the first buckling, whereas other models do not exhibit such a transitional phase. In addition, this inhibits further growth of the bar, consequently delaying/preventing the occurrence of the secondary
buckling instability.

The density slices at the $z=0$ and $y=0$ planes of model DM1 are displayed in the last row of Figures \ref{fig10face} and \ref{fig11side}, respectively. Until the end of the evolution, the mass inflow rate in model MD1 is about an order of magnitude smaller than that of model V1, $dM_{\star}(R<1R_d)/dt \approx1.06\times10^6\Msun \Gyr^{-1}$ after the buckling instability, resulting in no visible change in the face-on views. Without recurrent buckling instability, its morphology remains nearly the same after the first buckling instability, and thus the outer regions at $R\sim2.4\kpc$ are not thickened vertically.

\subsection{Stellar Mass}\label{s:ch4mass}

Model M1 with $M_d=1.4\times 10^{9} \Msun$ and model M2 with $0.8\times 10^{9} \Msun$ are designed to cover the observed range of stellar mass in VCC856 \citep{toloba14}. As shown in Figure \ref{fig01dynamics}, by adjusting the number of counter rotating particles, we make the mean streaming velocity $\vphi$ in these models similar to the observed rotation curve. We also assign appropriate halo concentrations in M models to keep $f_{\rm DM}$ the same as in model S1. This results in the almost identical $Q_T$ profiles (Fig.~\ref{fig02toomre}). These M models allow us to examine the combination of stabilizing/destabilizing effects of varying disk mass, mean streaming velocity, and halo concentration.

Figure \ref{fig17dmm} also plots the temporal variations of $F_2/F_0$, $L_z$, and $\Omega_b$ of models M1 and M2 in comparison with model S1.
The bar strength in M models starts to increase almost at the same time as in model S1. Although model M1 forms the bar slightly more strongly and undergoes buckling instability earlier than the others, the models' overall evolutionary paths are not significantly different. The last two rows of Figure \ref{fig16h2} show that model M1 forms a longer bar that is more efficient in transferring angular momentum than models with lower $M_d$, resulting in a slightly faster bar growth. This evidently leads to a earlier and slightly more vigorous buckling instability than in model M2, while the second buckling instability is more prominent in the latter (see Section \ref{s:ch4v}). Although the bar pattern speed is initially larger in a more massive disk, it becomes almost independent of the disk mass at 10 Gyr due to a larger rate of angular momentum transfer.

Figure \ref{fig17dmm}(a) shows that model M2 is more unstable and undergoes the first buckling episode slightly earlier than model S1, even though the former has a disk 1.25 times less massive than the latter. This is primarily because the disk in model M2 has a smaller fraction of counter streaming particles and its destabilizing overcomes the stabilizing effect of a low disk mass.

\section{Summary and Discussions}\label{s:ch5}

In this study, we have presented the results of $N$-body simulations to study the stability of dwarf disk galaxies. We initially set up galaxy models that resemble VCC856, a real galaxy in the Virgo cluster, assuming it is an infalling progenitor of dEdis. This is the first numerical study that investigates the stability and dynamical evolution of dwarf disk galaxies in isolation. We have also examined the effects of various parameters, such as counter streaming motion, velocity dispersion, disk scale height, dark matter fraction, and disk mass on the bar formation, buckling instability, regrowth of a bar, within the observational error ranges of VCC856. The main results of the present work can be summarized as follows.

We find that most of our galaxy models are dynamically unstable to forming a bar.
Only when galaxies have an excessively concentrated halo or a hot disk\footnote{The two stable models DM2 and DP4 initially contain a fraction of disk stars in counter streaming motions with a relatively thicker disk than normal disk galaxies, which causes an additional stabilizing effect.} do they remain stable without forming a bar. This suggests that infalling dwarf disk galaxies are \textit{intrinsically unstable} to bar formation even without tidal forcing. Our results explain the presence of dEdis with a bar component at the outskirts of Virgo cluster \citep{janz12,janz14} where the tidal force by the cluster potential appears too weak to produce a bar \citep{lokas16}.

The mean formation time before the first buckling is found to be $\sim6$ Gyr, indicating that the progenitors of dEdis slowly form a bar.
Except for two highly unstable models that have no counter streaming motion or low $f_{\rm{DM}}$, the bar formation in our dEdis models does not occur rapidly during their evolution in isolation. Our standard model is marginally unstable to forming a bar after $t\sim2.0\Gyr$. This may allow feasible time for the formation of very weak spiral arms in VCC 856 via either swing amplification or weak tidal force from the neighboring galaxies, as suggested by \cite{jerjen00}. We defer further discussion regarding the spiral structure of VCC 856 and dEdis to the next paper of this series.

The bar formation time and its strength depend on the galaxy properties: a bar is stronger and forms earlier in galaxies with lower fraction of counter-streaming disk particles, lower halo concentration, lower velocity anisotropy $f_R=(\sigma_{R}/\sigma_{z})^2$, smaller scale height of a disk, and/or more massive disk. In particular, the ratio of the velocity dispersions, $\sigma_{R}/\sigma_{z}$, is directly related to the disk stability and the onset of buckling instability. A disk with a higher $\sigma_{R}$ is initially  more stable to bar formation and can be subject to the buckling instability even without forming a bar since $\sigma_{R}/\sigma_{z}$ is closer to the threshold value. Consistently with the results of \cite{athanassoula86} and \cite{athanassoula03}, a bar formed in models with larger $\sigma_{R}/\sigma_{z}$  is thus wider, weaker, and inefficient in transferring the bar angular momentum to the halo.

Interestingly, we also find that, despite variations in disk mass, the disk-halo systems with nearly the same $\vphi$ and $f_{\rm{DM}}$ within $R_{\rm{eff}}$ undergo a similar evolutionary path. For instance, a more massive disk that is more unstable can be stabilized by increasing the fraction of counter-streaming particles in order to keep $\vphi$ and $f_{\rm{DM}}$ more or less the same. Such a combination of the destabilizing effect of disk mass and the stabilizing effect of counter-streaming motions makes the bar formation time in model M1 with a massive disk almost identical to that in model M2 with a less massive disk.

\begin{figure*}
\centering\includegraphics[angle=0,width=17cm]{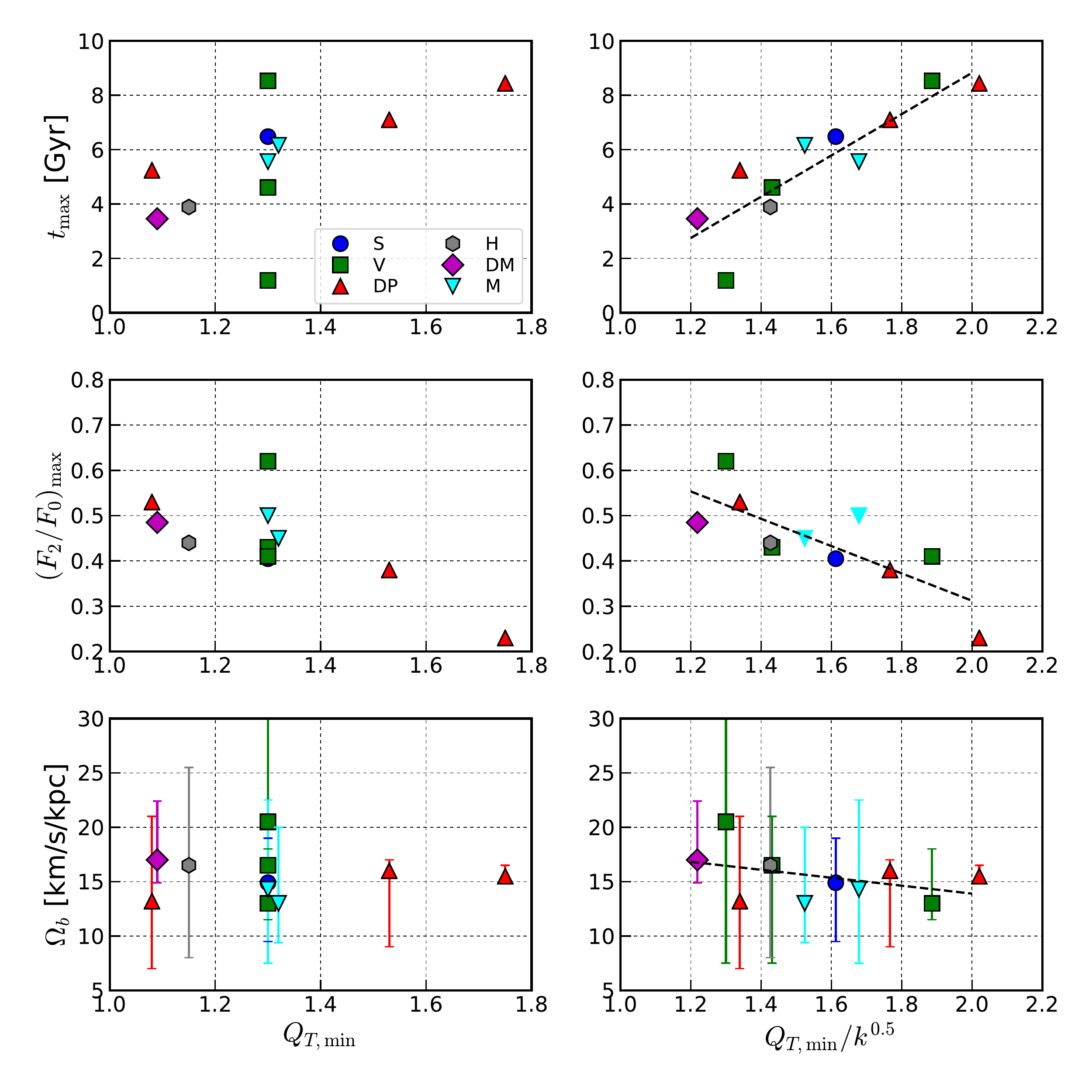}
\caption{Dependence of (top) the time $t_{\rm max}$ when the bar becomes strongest before the first buckling instability, (middle) the bar strength $(F_2/F_0)_{\rm max}$ at $t=t_{\rm max}$, and (bottom) the bar pattern speed $\Omega_b$ upon (left) $Q_{T,\rm min}$ and (right) $Q_{T,\rm min}/k^{0.5}$.  Only the models undergoing the first buckling instability are shown. In the bottom panels, the error bars mark the full ranges of the pattern speed over the course of evolution up to 10 Gyr, while the symbols give $\Omega_b$ at $t=t_{\rm max}$. The dashed lines in the right panels are the best fits whose forms are given in the text.}\label{fig18sum}
\end{figure*}

As predicted by linear theory (e.g., \citealt{kato71, lyndenbell72, tremaine84, weinberg85, athanassoula03}), the exchange of angular momentum between the disk and the halo plays an important role in shaping the dynamical properties of our galaxy models such as the bar growth and pattern speed. As a bar loses its angular momentum to the halo, it becomes stronger and its pattern speed decreases. Consequently, the bar growth rate and its slowdown rate are larger in models with lower velocity anisotropy $f_{R}$ or less counter streaming motions. Additionally, we find that thinner and more massive disks are more unstable and form a stronger bar, which efficiently transfers the angular momentum and thus rapidly slows down at least until the onset of buckling instability, if the surrounding halo is sufficiently massive. Model DM1 with $f_{\rm DM}=7\%$ is highly unstable to rapidly form a bar: however, due to a small amount of angular momentum transfer, it takes longer to undergo buckling instability relative to model V1.

The left panels of Figure \ref{fig18sum} summarize our numerical results for the time $t_{\rm max}$ when the bar is maximized before the first buckling instability, the maximum bar strength $(F_2/F_0)_{\rm max}$, and the bar pattern speed as functions of the minimum value $Q_{T,\rm min}$ of the Toomre stability parameter for the models with the buckling instability. Except for V models, a model with smaller $Q_{T,\rm min}$ forms a stronger bar at an earlier time.  We find that the degeneracy of $Q_{T,\rm min}$ on $t_{\rm max}$ and $(F_2/F_0)_{\rm max}$ for V models can easily be removed  by considering the amount of counter-streaming motion in the abscissa as $Q_{T,\rm min}/k^{0.5}$, as plotted in the right panels. The dashed lines are the best fits: $t_{\rm max}/{\rm Gyr} = -6.4 + 7.6 Q_{T,\rm min}/k^{0.5}$, $(F_2/F_0)_{\rm max}=0.9-0.3 Q_{T,\rm min}/k^{0.5}$, and $\Omega_b/({\rm km\,s^{-1}\,kpc^{-1}})=21.1-3.7Q_{T,\rm min}/k^{0.5}$.

The buckling instability appears to be quite common at least in barred dEdis, as it occurs in 11 of our 15 galaxy models. It shortens and thickens a bar by instantaneously warping it in a U or V shape when the velocity anisotropy $\sigma_z/\sigma_R$ exceeds a threshold value at a given radius, as in normal disk galaxies \citep{combes81,combes90,raha91,merrifield94,martinez04}.
In the V and DP models with the same density structure, the threshold value is
$\sigma_z/\sigma_R\approx0.63$ at $R=0.9\kpc$, which is similar to the value reported by \citet{martinez06} for normal disk galaxies. The buckling instability not only more vigorously dissolves the bar in a more unstable disk, but also causes a transition in $\lzdt$ and $d\Omega/dt$  (e.g., see model DP1 and H1 in Figs.~\ref{fig13dp1} and \ref{fig15h1}).

In a study of the effects of disk thickness on the evolution of barred galaxies, \cite{klypin09} found that a thinner disk forms a shorter bar and experiences buckling instability earlier than a thicker counterpart.
They argued that the disk scale height is related to the phase-space density as $f\equiv \rho_\star/(\sigma_{R}\sigma_{\phi}\sigma_{z})\propto z_{d}^{-3/2}$ and to the Jeans mass as $M_{J}\propto \sigma_{R}\sigma_{\phi}\sigma_{z}/\sqrt{\rho_\star}\propto z_{d}$, making a disk with smaller $z_d$ more unstable. These results are in good agreement with ours in terms of the epochs of bar formation and buckling instability, but are seemingly contradictory in terms of a tendency of bar slowdown. In \citet{klypin09}, a thin-disk bar ($K_{a2}$) barely slows down, resulting in a 2.5 times larger pattern speed than the thicker counterpart ($K_{a3}$) at $t=5$ Gyr, whereas, in our study, the thin-disk bar (model H1) slows down more significantly than a thicker counterpart (model S1). We attribute this inconsistency to the early onset of the buckling instability in the thin-disk bar of \citet{klypin09}. More specifically, their thin-disk model with $z_d/R_d=0.05$ is highly unstable to undergoing vigorous buckling instability before the bar grows full and long enough for an efficient angular momentum transfer.\footnote{Before the onset of the buckling instability, the bar strengths are 0.40 and 0.55 for their thin- and thick-disk bars.} Our thin-disk model H1 with $z_d/R_d=0.23$ is even thicker than their thickest-disk model $K_{a3}$ with $z_d/R_d=0.18$, and the evolutions of these two models are similar to one another in terms of the buckling time and bar slowdown.

While we study the evolution of dwarf disk galaxies, some of our results can be applicable to normal disk galaxies, at least qualitatively. For instance, model V1 presented in Section \ref{s:ch4v} follows an evolutionary path similar to a Milky Way-sized disk galaxy in \cite{martinez06}. Both undergo more than one buckling instability, and the times of the recurrent buckling instabilities are similar. The ellipticity variations are also similar (see Fig.~\ref{fig09v2}(a) and (b) in the present work and their Fig.~4).
In addition, the face-on view at 7 Gyr of model V1 (Fig.~\ref{fig10face}) closely resembles the image with an ansae at 9.9 Gyr (their Fig.~11). However, unlike the bar in model V1, the bar in their model keeps growing in size  after the second buckling instability. Other than this minor difference, their evolution histories are almost alike.

The formation of barred spirals and inner stellar rings observed in Milky Way-sized galaxies \citep{athanassoula02,athanassoula12,lokas16a} is also found in model V1 without counter-streaming motions (see Fig.~\ref{fig12ring}).
Model V1 is very unstable to rapidly forming a bar and transfers a significant amount of bar angular momentum to the halo. This allows the bar to grow longer than others to beyond the CR, triggering the ejection of stars at the bar ends and forming spiral patterns in a trailing shape \citep{athanassoula12, lokas16a}. The angular momentum transfer occurs even after the buckling instability, which eventually aids the formation of an inner stellar ring.
As we find that the increase in the central enclosed mass is correlated with the angular momentum transfer, the mass inflow rate at $1R_d$ in model V1 is 1.6 times higher than our standard model during the post-buckling era. This allows the cavities at the sides of the bar to grow larger to transform the disk into a $\Theta$ shape.

A bar formed in dwarf disk galaxies is subject to multiple buckling instabilities that vertically heat the disk. The first buckling episode that occurs at relatively smaller radii (1$\sim$3$R_{d}$) often forms a peanut bulge and vertically stabilizes the central regions. On the other hand, the secondary bucklings occur at relatively larger radii ($R>3R_{d}$), so only the outer parts are thickened by the vertical heating, forming an X-shaped bulge.
This is consistent with \cite{martinez06}, who investigated the recurrent buckling instability in Milky Way-sized galaxies and suggested that an X-shaped bulge is evidence of recurrent buckling. Models where the secondary buckling is relatively mild form a faint X-shaped bulge (e.g., model H1 and DP1).

Among our 15 realistic dwarf disk galaxies, 13 models form bars and 9 experience recurrent buckling instabilities. This suggests that bar regrowth and ensuing recurrent bucklings are quite common, at least for the progenitors of dEdis. The regrowth is not observed in model DM1 with small $f_{\rm DM}$ due to the lack of the angular momentum exchange. This is consistent with \cite{martinez06}, who showed that the regrowth of a bar is mediated by the resonant interaction between the bar and the surrounding halo. Our numerical results suggest that not only the disk-halo ratio, but also the strength of the first buckling, is essential for the regrowth. Between models V1 and V2, for example, the first buckling instability in model V1 is sufficiently strong to significantly dissolve the bar, leaving only a small number of stars near the resonant bar-end regions for the regrowth. Consequently, regrowth occurs faster in model V2 and the resulting the secondary buckling is stronger than in model V1.

\acknowledgments
We are grateful to the referee for the valuable comments that helped us considerably improve the paper. We gratefully acknowledge Thomas Quinn for helpful comments on N-body simulations using  the \textsc{changa} code. We also thank Volker Springel and Denis Yurin for the helpful discussions on constructing initial conditions using the \textsc{galic} code. S.K. was supported by the Brain Korea 21 program and In-Ha Kim scholarship. S.C.R. acknowledges support from the Basic Research Program through the NRF funded by the Ministry of Education, Science, and Technology (2015R1A2A2A01006828) and the Center for Galaxy Evolution Research (No.~2010-0027910). This work was supported by the National Research Foundation of Korea (NRF) grant funded by the Korean government (MEST) (No.~3348-20160021). The computation of this work was supported by the Supercomputing Center/Korea Institute of Science and Technology Information with supercomputing resources including technical support (KSC-2015-C3-027).

\end{document}